\pdfoutput=1

\documentclass[12pt,a4paper]{article}

\usepackage{ifthen} % for conditional statements
\newboolean{pdflatex}
\setboolean{pdflatex}{true} % False for eps figures 

\usepackage{titlesec}

\titleformat*{\section}{\large\bfseries}
\titleformat*{\subsection}{\large\bfseries}

\newboolean{articletitles}
\setboolean{articletitles}{true} % False removes titles in references

\newboolean{uprightparticles}
\setboolean{uprightparticles}{false} %True for upright particle symbols

\def\paperauthors{Stephen Farry$^1$, Olli Lupton$^2$, Martina Pili$^3$, Mika Vesterinen$^2$}
\def\paperasciititle{Understanding and constraining the PDF uncertainties in a W boson mass measurement with forward muons at the LHC} % Set ASCII title here
\def\papertitle{\LARGE{Understanding and constraining the PDF uncertainties in a $W$ boson mass measurement with forward muons\\ at the LHC}} % Latex formatted title 
\date{\vspace{-8ex}}
\def\paperkeywords{{High Energy Physics}, {LHCb}} % Comma separated list
\def\papercopyright{CERN on behalf of the LHCb collaboration}

\def\paperlicenceurl{https://creativecommons.org/licenses/by/4.0/}

\usepackage[top=1in, bottom=1.25in, left=1in, right=1in]{geometry}

\usepackage{microtype}
\usepackage{lineno}  % for line numbering during review
\usepackage{xspace} % To avoid problems with missing or double spaces after
\usepackage{caption} %these three command get the figure and table captions automatically small

\usepackage{subfig}
\usepackage{graphicx}  % to include figures (can also use other packages)
\usepackage{color}
\usepackage{colortbl}
\graphicspath{{./figs/}} % Make Latex search fig subdir for figures

\usepackage{amsmath} % Adds a large collection of math symbols
\usepackage{amssymb}
\usepackage{amsfonts}
\usepackage{upgreek} % Adds in support for greek letters in roman typeset

\newcommand*\patchAmsMathEnvironmentForLineno[1]{%
\expandafter\let\csname old#1\expandafter\endcsname\csname #1\endcsname
\expandafter\let\csname oldend#1\expandafter\endcsname\csname
end#1\endcsname
 \renewenvironment{#1}%
   {\linenomath\csname old#1\endcsname}%
   {\csname oldend#1\endcsname\endlinenomath}%
}
\newcommand*\patchBothAmsMathEnvironmentsForLineno[1]{%
  \patchAmsMathEnvironmentForLineno{#1}%
  \patchAmsMathEnvironmentForLineno{#1*}%
}
\AtBeginDocument{%
\patchBothAmsMathEnvironmentsForLineno{equation}%
\patchBothAmsMathEnvironmentsForLineno{align}%
\patchBothAmsMathEnvironmentsForLineno{flalign}%
\patchBothAmsMathEnvironmentsForLineno{alignat}%
\patchBothAmsMathEnvironmentsForLineno{gather}%
\patchBothAmsMathEnvironmentsForLineno{multline}%
\patchBothAmsMathEnvironmentsForLineno{eqnarray}%
}

\usepackage{hyperxmp}

\usepackage[pdftex,
            pdfauthor={\paperauthors},
            pdftitle={\paperasciititle},
            pdfkeywords={\paperkeywords},
            pdfcopyright={Copyright (C) \papercopyright},
            pdflicenseurl={\paperlicenceurl}]{hyperref}

\usepackage[all]{hypcap} % Internal hyperlinks to floats.

\usepackage{xspace} 
\usepackage{upgreek}

\def\MagUp {\mbox{\em Mag\kern -0.05em Up}\xspace}

\ifthenelse{\boolean{uprightparticles}}%
{

 \def\PDelta      {\ensuremath{\Delta}\xspace}                 
 \def\PXi      {\ensuremath{\Xi}\xspace}                 
 \def\PLambda      {\ensuremath{\Lambda}\xspace}                 
 \def\PSigma      {\ensuremath{\Sigma}\xspace}                 
 \def\POmega      {\ensuremath{\Omega}\xspace}                 
 \def\PUpsilon      {\ensuremath{\Upsilon}\xspace}                 
 
 \def\PB      {\ensuremath{\mathrm{B}}\xspace}                 
                  
 \def\PD      {\ensuremath{\mathrm{D}}\xspace}

 \def\PK      {\ensuremath{\mathrm{K}}\xspace}

 \def\Pi      {\ensuremath{\mathrm{i}}\xspace}

}
{

 \mathchardef\PDelta="7101
 \mathchardef\PXi="7104
 \mathchardef\PLambda="7103
 \mathchardef\PSigma="7106
 \mathchardef\POmega="710A
 \mathchardef\PUpsilon="7107
                  
 \def\PB      {\ensuremath{B}\xspace}                 
                  
 \def\PD      {\ensuremath{D}\xspace}

 \def\PK      {\ensuremath{K}\xspace}

 \def\Pi      {\ensuremath{i}\xspace}

}

\makeatletter
\ifcase \@ptsize \relax% 10pt
  \newcommand{\miniscule}{\@setfontsize\miniscule{4}{5}}% \tiny: 5/6
\or% 11pt
  \newcommand{\miniscule}{\@setfontsize\miniscule{5}{6}}% \tiny: 6/7
\or% 12pt
  \newcommand{\miniscule}{\@setfontsize\miniscule{5}{6}}% \tiny: 6/7
\fi
\makeatother

\DeclareRobustCommand{\optbar}[1]{\shortstack{{\miniscule (\rule[.5ex]{1.25em}{.18mm})}
  \\ [-.7ex] $#1$}}

  \def\Kbar    {{\kern 0.2em\overline{\kern -0.2em \PK}{}}\xspace}

\def\KorKbar    {\kern 0.18em\optbar{\kern -0.18em K}{}\xspace}

  \def\Dbar    {{\kern 0.2em\overline{\kern -0.2em \PD}{}}\xspace}

\def\DorDbar    {\kern 0.18em\optbar{\kern -0.18em D}{}\xspace}

\def\Bbar    {{\ensuremath{\kern 0.18em\overline{\kern -0.18em \PB}{}}}\xspace}

\def\BorBbar    {\kern 0.18em\optbar{\kern -0.18em B}{}\xspace}

  \def\Y#1S{\ensuremath{\PUpsilon{(#1S)}}\xspace}% no space before {...}!

\def\Lbar        {{\ensuremath{\kern 0.1em\overline{\kern -0.1em\PLambda}}}\xspace}
\def\LorLbar    {\kern 0.18em\optbar{\kern -0.18em \PLambda}{}\xspace}

\def\to                 {\ensuremath{\rightarrow}\xspace}

\def\AT#1     {\ensuremath{A_{\mathrm{T}}^{#1}}\xspace}           % 2

\def\C#1      {\ensuremath{\mathcal{C}_{#1}}\xspace}                       % 9
\def\Cp#1     {\ensuremath{\mathcal{C}_{#1}^{'}}\xspace}                    % 7
\def\Ceff#1   {\ensuremath{\mathcal{C}_{#1}^{\mathrm{(eff)}}}\xspace}        % 9  
\def\Cpeff#1  {\ensuremath{\mathcal{C}_{#1}^{'\mathrm{(eff)}}}\xspace}       % 7
\def\Ope#1    {\ensuremath{\mathcal{O}_{#1}}\xspace}                       % 2
\def\Opep#1   {\ensuremath{\mathcal{O}_{#1}^{'}}\xspace}                    % 7

\newcommand{\tev}{\ifthenelse{\boolean{inbibliography}}{\ensuremath{~T\kern -0.05em eV}}{\ensuremath{\mathrm{\,Te\kern -0.1em V}}}\xspace}
\newcommand{\gev}{\ensuremath{\mathrm{\,Ge\kern -0.1em V}}\xspace}
\newcommand{\mev}{\ensuremath{\mathrm{\,Me\kern -0.1em V}}\xspace}
\newcommand{\kev}{\ensuremath{\mathrm{\,ke\kern -0.1em V}}\xspace}
\newcommand{\ev}{\ensuremath{\mathrm{\,e\kern -0.1em V}}\xspace}
\newcommand{\gevc}{\ensuremath{{\mathrm{\,Ge\kern -0.1em V\!/}c}}\xspace}
\newcommand{\mevc}{\ensuremath{{\mathrm{\,Me\kern -0.1em V\!/}c}}\xspace}
\newcommand{\gevcc}{\ensuremath{{\mathrm{\,Ge\kern -0.1em V\!/}c^2}}\xspace}
\newcommand{\gevgevcccc}{\ensuremath{{\mathrm{\,Ge\kern -0.1em V^2\!/}c^4}}\xspace}
\newcommand{\mevcc}{\ensuremath{{\mathrm{\,Me\kern -0.1em V\!/}c^2}}\xspace}

\def\gsim{{~\raise.15em\hbox{$>$}\kern-.85em
          \lower.35em\hbox{$\sim$}~}\xspace}
\def\lsim{{~\raise.15em\hbox{$<$}\kern-.85em
          \lower.35em\hbox{$\sim$}~}\xspace}

\def\tell1  {TELL1\xspace}
\def\ukl1   {UKL1\xspace}

\usepackage{cite} % Allows for ranges in citations
\usepackage{mciteplus}

\usepackage{longtable}
\usepackage{float}
\usepackage{multicol}
\usepackage{booktabs}
\usepackage{placeins}

\begin{document}
%%%%%%%%%%%%%%%%%%%%%%%%%
%%%%% Title     %%%%%%%%%
%%%%%%%%%%%%%%%%%%%%%%%%%
\renewcommand{\thefootnote}{\fnsymbol{footnote}}

\setcounter{footnote}{1}

% %%%%%%%%%%%%% TITLE PAGE--------
% $Id: title-LHCb-ANA.tex 39841 2013-07-26 10:31:08Z roldeman $
% ===============================================================================
% Purpose: LHCb-ANA Note title page template
% Author: 
% Created on: 2010-10-05
% ===============================================================================

%%%%%%%%%%%%%%%%%%%%%%%%%
%%%%%  TITLE PAGE  %%%%%%
%%%%%%%%%%%%%%%%%%%%%%%%%
\begin{titlepage}

% Header ---------------------------------------------------
\vspace*{-1.5cm}

\noindent
\begin{tabular*}{\linewidth}{lcr}
\\
&& \\ 
%Review - Version 10  & & \\  % ID 
% \today & &  \\ % Date - Can also hardwire e.g.: 23 March 2010
 & & \\
\hline
\end{tabular*}

\vspace*{4.0cm}

% Title --------------------------------------------------
{\normalfont\bfseries\boldmath\huge
\begin{center}
% DO NOT EDIT HERE. Instead edit macro in main.tex to keep metadata correct
  \papertitle
\end{center}
}

\vspace*{2.0cm}

% Authors -------------------------------------------------
\begin{center}
% If changing to list here, make pdfauthors in main.tex a comma
% separated list with the same names. Otherwise metadata in file will be wrong.
\paperauthors.
\bigskip\\
{\normalfont\itshape\footnotesize
$ ^1$University of Liverpool, Liverpool, United Kingdom\\
$ ^2$University of Warwick, Coventry, United Kingdom\\
%$ ^2$European Organization for Nuclear Research (CERN), Geneva, Switzerland\\
$ ^3$University of Oxford, Oxford, United Kingdom\\

}
\end{center}

\vspace{\fill}

% Abstract -----------------------------------------------
\begin{abstract}
  \noindent
Precision electroweak tests are a powerful probe of physics beyond the Standard Model, but the sensitivity is limited by the precision with which the $W$ boson mass ($M_W$) has been measured. The Parton Distribution Function (PDF) uncertainties are a potential limitation for measurements of $M_W$ with LHC data.
It has recently been pointed out that, thanks to LHCb's unique forward rapidity acceptance, a new measurement of $M_W$ by LHCb can improve this situation. 
%While the measurement of $M_W$ by LHCb would also be susceptible to uncertainties in the PDFs, there would be a partial cancellation of the overall PDF uncertainty 
%when the LHCb result is included in an average with measurements by ATLAS and CMS. 
Here we report on a detailed study on the mechanism driving the PDF uncertainty in the LHCb measurement of $M_W$, 
and propose an approach which should reduce this uncertainty by roughly a factor of two using LHCb Run 2 data.
\end{abstract}

\vspace*{2.0cm}
\vspace{\fill}

\end{titlepage}

\pagestyle{empty}  % no page number for the title 

%%%%%%%%%%%%%%%%%%%%%%%%%%%%%%%%
%%%%%  EOD OF TITLE PAGE  %%%%%%
%%%%%%%%%%%%%%%%%%%%%%%%%%%%%%%%

% %%%%%%%%%%%%% ---------

\renewcommand{\thefootnote}{\arabic{footnote}}
\setcounter{footnote}{0}
\tableofcontents

%%%%%%%%%%%%%%%%%%%%%%%%%
%%%%% Main text %%%%%%%%%
%%%%%%%%%%%%%%%%%%%%%%%%%

\pagestyle{plain} % restore page numbers for the main text
\pagenumbering{arabic}

\section{Introduction}
\label{sec:Introduction}
Global fits to precision electroweak data are sensitive to physics beyond the Standard Model (SM). Of notable interest is the mass of the $W$ boson ($M_W$)
because, currently, it is predicted with higher precision than it is measured.
The 2018 update of the electroweak fit by the gFitter collaboration indirectly predicts $M_W = 80354 \pm 7$\,MeV/c$^2$~\cite{Gfitter}. 
This prediction is more precise than the average of direct measurements reported by the Particle Data Group, $M_W = 80379 \pm 12$\,MeV/c$^2$~\cite{PDG}, which is dominated by measurements using $W\rightarrow \ell \nu_\ell$ decays at hadron collider experiments, where $\ell$ can be either an electron or a muon.  

Measurements of $M_W$ at hadron colliders are performed by comparing data to templates of the charged lepton transverse momentum, missing transverse energy, and transverse mass in samples of $W\rightarrow \ell \nu_\ell$ decays. The combination of measurements by the CDF~\cite{CDF1} and D0~\cite{D0} experiments at the Fermilab Tevatron $p\bar{p}$ collider is $M_W = 80387 \pm 16$\, MeV/c$^2$~\cite{Tevatron}. 
In $p\bar{p}$ collisions $W$ bosons are primarily produced by the annihilation of valence quarks and antiquarks. By contrast, gluons and sea quarks play a critical role in the $pp$ collisions at the LHC. Measurements of $M_W$ at the LHC are therefore expected to be more susceptible to theoretical uncertainties in the modeling of $W$ production, in particular those related to the Parton Distribution Functions (PDFs), than at the Tevatron~\cite{Krasny,Bozzi,Rojo,Vicini,Quackenbush}.  The ATLAS Collaboration reported a measurement of $M_W = 80370 \pm 13 \pm 14$\,MeV/c$^2$ where the first and second uncertainties are experimental and theoretical, respectively~\cite{ATLAS}. The dominant contribution to the theoretical uncertainty can be attributed to the PDFs. A key challenge of future measurements by ATLAS and CMS will be to reduce the PDF uncertainty. 
 
The current ATLAS and CMS detectors are capable of reconstructing charged leptons in the approximate pseudorapidity range $|\eta|<2.5$, where $\eta = -\ln (\tan(\theta/2))$ with $\theta$ being the angle between the particle direction and the beam axis.
LHCb~\cite{LHCb_det} is a single-arm spectrometer with full charged particle tracking and identification capabilities over the range $2 < \eta < 5$, which is mostly orthogonal to the acceptance of ATLAS and CMS. While LHCb is primarily designed for the study of beauty and charm hadrons, it has a strong track record in measurements of $W$ and $Z$ production in muonic final states~\cite{WZ8,Z13}. As for precision electroweak tests, LHCb has already measured the effective weak mixing angle $\sin^2\theta_{\rm eff}^{\rm lept}$~\cite{sin2theta}, but the potential for a measurement of $M_W$ was not realised until recently.
 
\begin{figure}
\centerline{
\includegraphics[width = 0.5\textwidth]{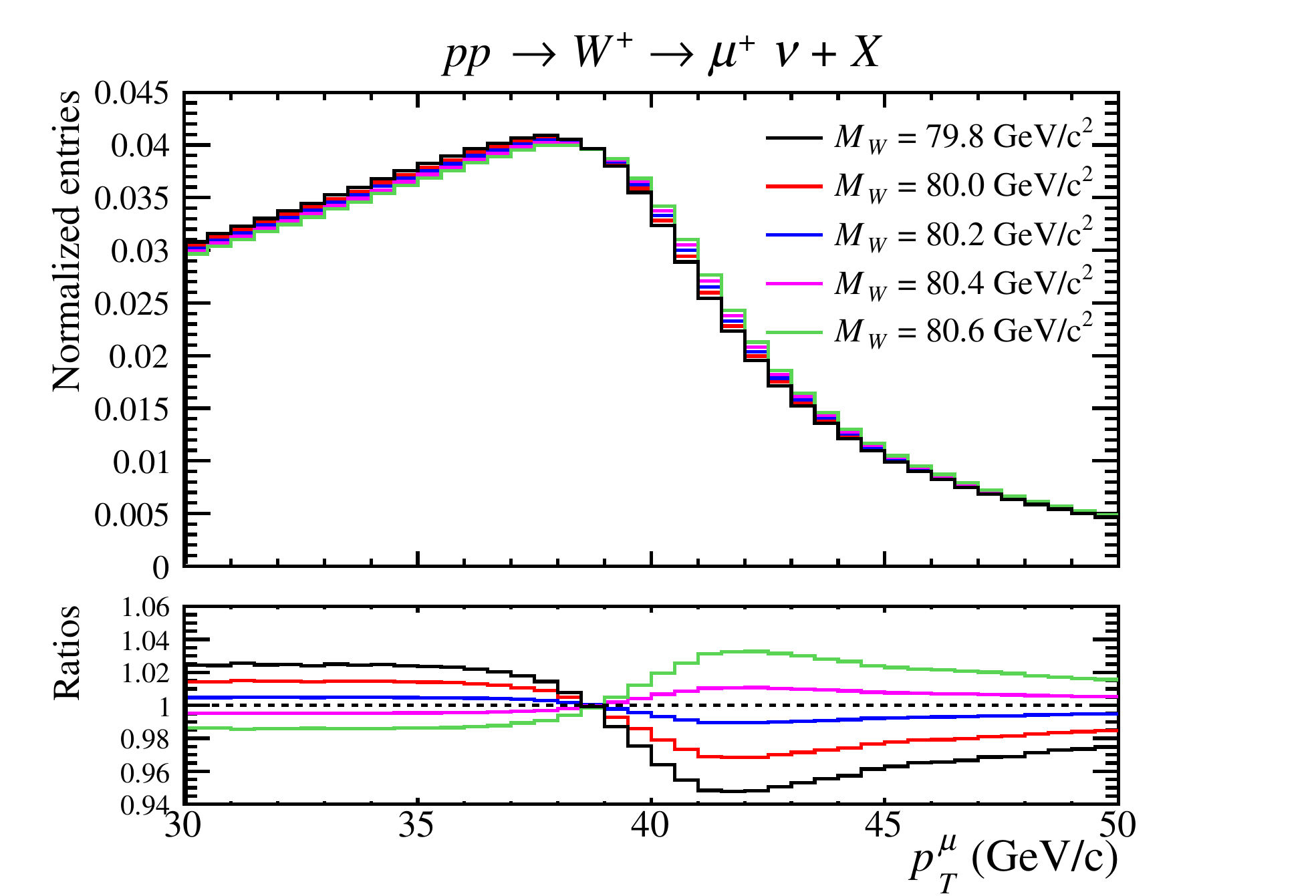}\includegraphics[width = 0.5\textwidth]{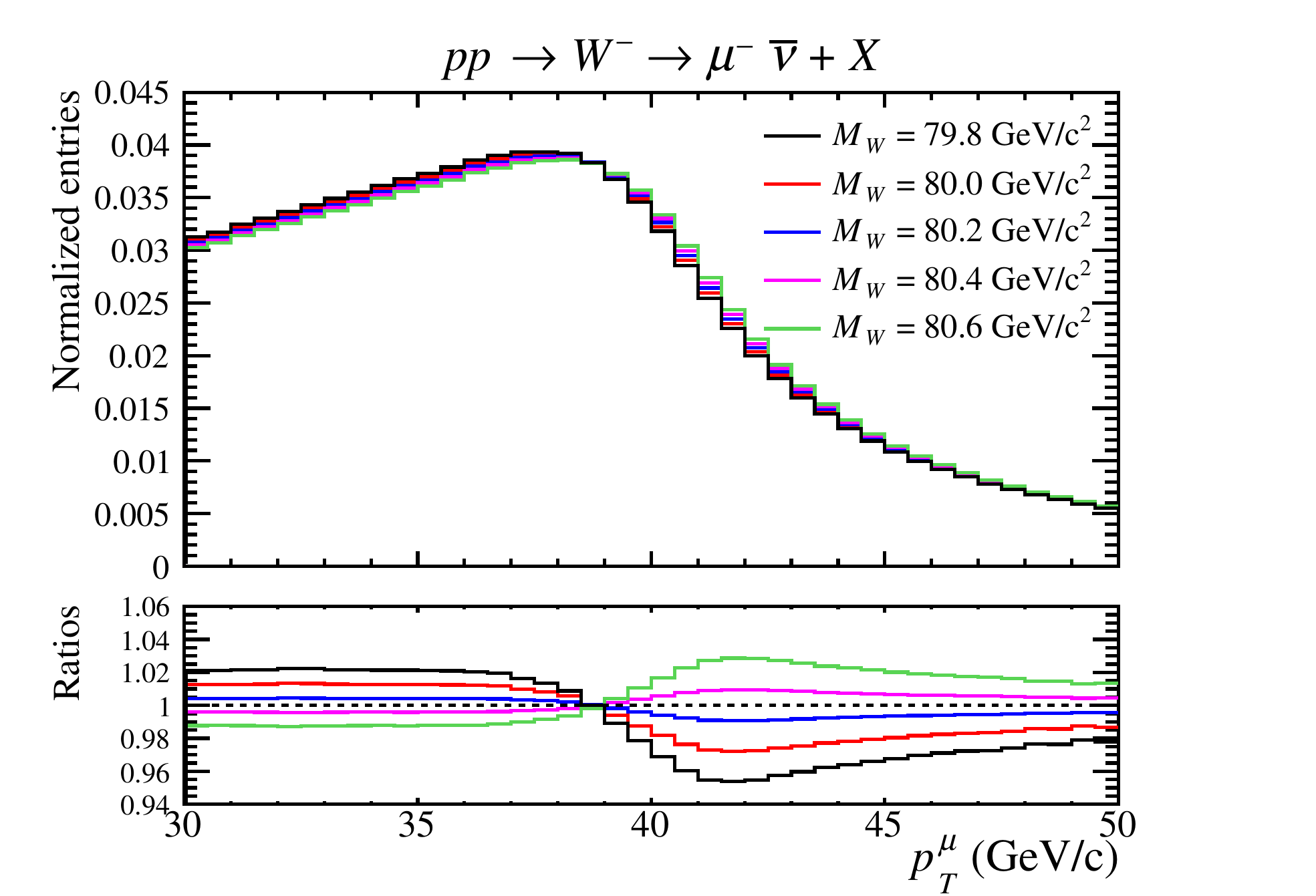}}
\caption{The simulated muon $p_T^\mu$ distributions in $W\rightarrow \mu\nu$ decays (left $W^+$, right $W^-$) with five different $M_W$ hypotheses. The ratios are with respect to the prediction with $M_W = 80.3$\,GeV/c$^2$.} 
\label{fig:templates}
\end{figure}

Ref.~\cite{LHCb} proposed a new measurement of $M_W$ by LHCb based on the muon transverse momentum ($p_T^\mu$) distribution with $W \to \mu \nu$ decays. 
Fig.~\ref{fig:templates} shows how the shape of the $p_T^\mu$ distribution varies with the $M_W$ hypothesis in simulated events. The maximum variation in the normalised distribution, which occurs at $p_T^\mu \sim$ 42\,GeV/c, is around $10^{-4}$ per MeV/c$^2$ of shift in $M_W$. Large $W$ samples are therefore required to resolve this subtle change in the shape of the $p_T^\mu$ distribution. After the successful completion of LHCb Run 2 roughly 6\,fb$^{-1}$ of $pp$ collisions at $\sqrt{s} =$ 13\,TeV have been recorded, complementing the 3\,fb$^{-1}$ recorded at lower $\sqrt{s}$ values in Run 1. Using the methods described in this paper we estimate that the Run 2 data could yield a $M_W$ measurement with a statistical uncertainty of roughly 10\,MeV/c$^2$. The obvious next question is how well the theoretical uncertainties, in particular those related to the PDFs, can be controlled.
Ref.~\cite{LHCb} estimated that the PDF uncertainties in a standalone LHCb measurement would be larger than those in ATLAS and CMS. However, the uncertainty on the LHCb measurement would be partially anticorrelated with those of ATLAS and CMS.  It is therefore claimed that the introduction of a LHCb measurement into a LHC $M_W$ average could reduce the overall PDF uncertainty. Similar improvements may be possible with the extended angular coverage of the upgraded ATLAS and CMS detectors in the HL-LHC era, as explored in a recent study by ATLAS~\cite{ATLAS_HL}. Given the large size of the LHCb Run 2 dataset, and anticipated future data with LHCb Upgrade I~\cite{LHCb_up1} and the proposed Upgrade II~\cite{LHCb_up2}, it seems worthwhile to study in greater detail the cause of the PDF uncertainty in a measurement of $M_W$ by LHCb, and possible strategies to reduce it. 
\section{Simulation of \boldmath{$W$} production}
\label{sec:simulation}
A sample of $10^{8}$ Monte Carlo events of the 
type $pp\rightarrow W \rightarrow \mu\nu + X$, at a centre-of-mass energy $\sqrt{s}=$ 13\,TeV, is generated
using POWHEG~\cite{powheg} with the CT10~\cite{CT10} PDFs.
These events are subsequently processed with PYTHIA~\cite{pythia} to simulate the parton showering. 
No LHCb detector response is simulated. 
Unless otherwise specified, events are analysed if they satisfy $30<p_T^\mu<50$\,GeV/c and $2<|\eta|<4.5$\footnote{The $2<|\eta|<4.5$ selection is chosen to make better use of the available samples: the events falling in the negative $\eta$ region are equivalently treated as those with positive $\eta$.}. Roughly 10\% of the initial event sample falls into this kinematic region. 
The invariant mass of the $W$ decay products ($m$) is assumed to follow a relativistic Breit-Wigner distribution:
\begin{equation}
\frac{d\sigma}{dm} \propto \frac{m^2}{(m^2-M_W^2)^2+ m^4\Gamma_W^2/M_W^2},\label{eq:BW}
\end{equation}
where $M_W$ and $\Gamma_W$ are the mass and the width of the $W$ boson, respectively. The events are generated with a nominal value of $M_W$~\cite{PDG} but can be reweighted according to Eq.~\ref{eq:BW} to emulate a different $M_W$ hypothesis. 
%Fig.~\ref{fig:templates} shows how the shape of the $p_T$ spectrum is distorted by variations in $M_W$. 
 
A similar set of weights can be assigned to map the sample to different PDFs. 
As in Ref.~\cite{LHCb} the full PDF uncertainty should consider an envelope of PDF sets from several groups, including for example the MMHT14~\cite{MMHT} and CT14~\cite{CT14} sets, but for the current study 
we focus on the NNPDF3.1~\cite{NNPDF} set with 1000 equiprobable {\em replicas}. 
\section{Fitting method}
\label{sec:toy_method}
Scaling the generated event samples to the 6\,fb$^{-1}$ of LHCb Run 2 data yields an expectation of 7.2 (4.8) million $W^+$ ($W^-$) events in the 30 $<p_T^\mu<$ 50\,GeV/c and 2 $<\eta<$ 4.5 region. 
Toy data histograms are generated by randomly fluctuating the bins around the nominal distribution, assuming these yields and Poisson statistics. These histograms can be generated with different PDF sets using the reweighting procedure already described. The current study neglects experimental systematic uncertainties, such as those due to the knowledge of the momentum scale and the dependence of the muon identification efficiency on $p_T^\mu$ and $\eta$, 
%and does not address the systematic uncertainties related to the $p_T^W$ modelling~\cite{ptW1,Mangano}.
and does not address the treatment of higher order QCD corrections in the $p_T^W$ modelling~\cite{ptW1,Mangano}.

The data histograms are compared to templates with different PDF and $M_W$ hypotheses.
The normalisation of each template is scaled to match the data such that the fit only considers the shape information.
For a given PDF hypothesis a single-parameter (1D) fit determines the value of $M_W$ that minimises the $\chi^2$ between a toy and the templates.   
The 68\%~C.L.~statistical uncertainty corresponds to a variation of $\Delta\chi^2 = 1$ with respect to the parabola minimum.

\begin{figure}
 \centerline{   
\includegraphics[width=0.5\textwidth]{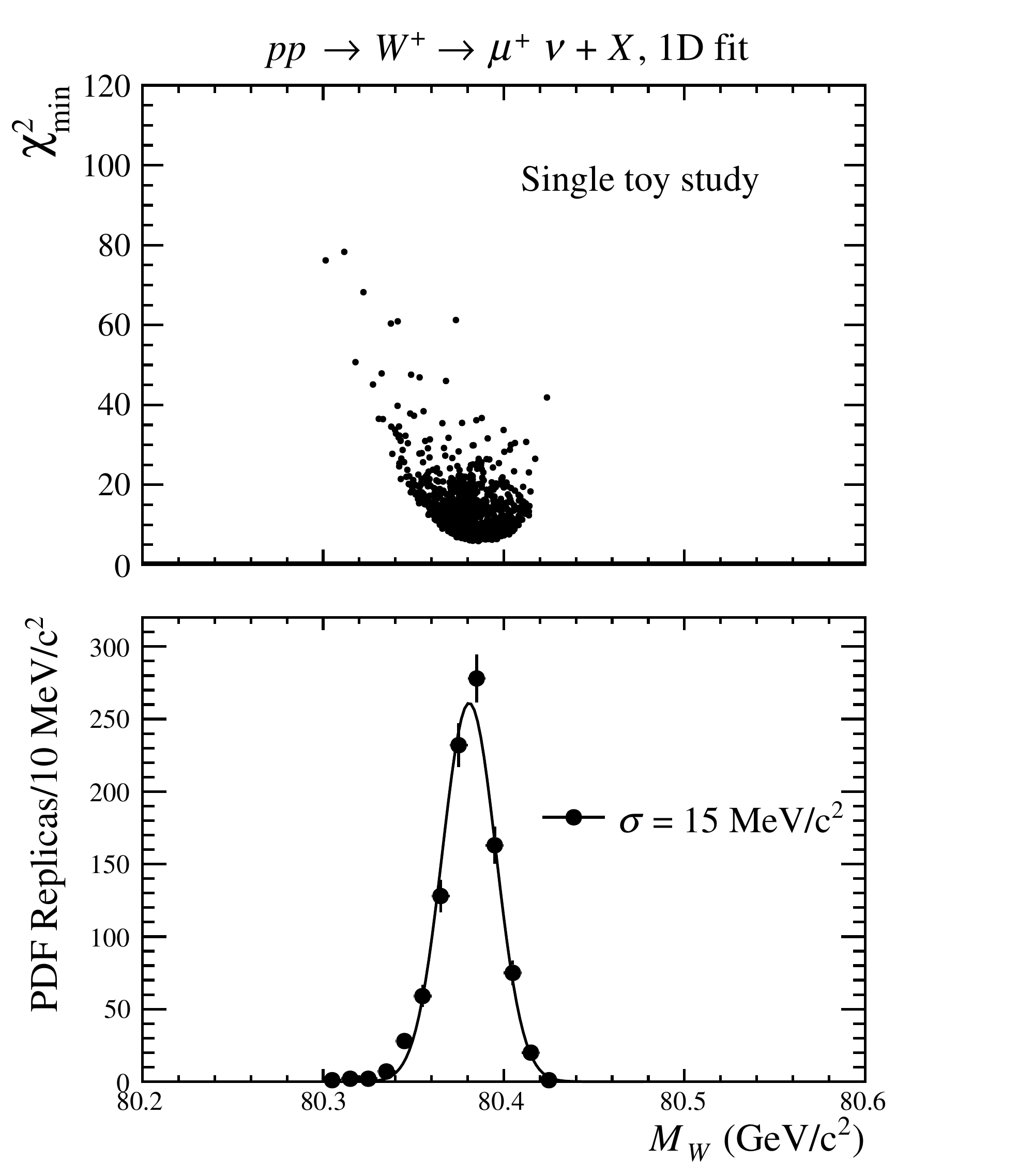}\includegraphics[width=0.5\textwidth]{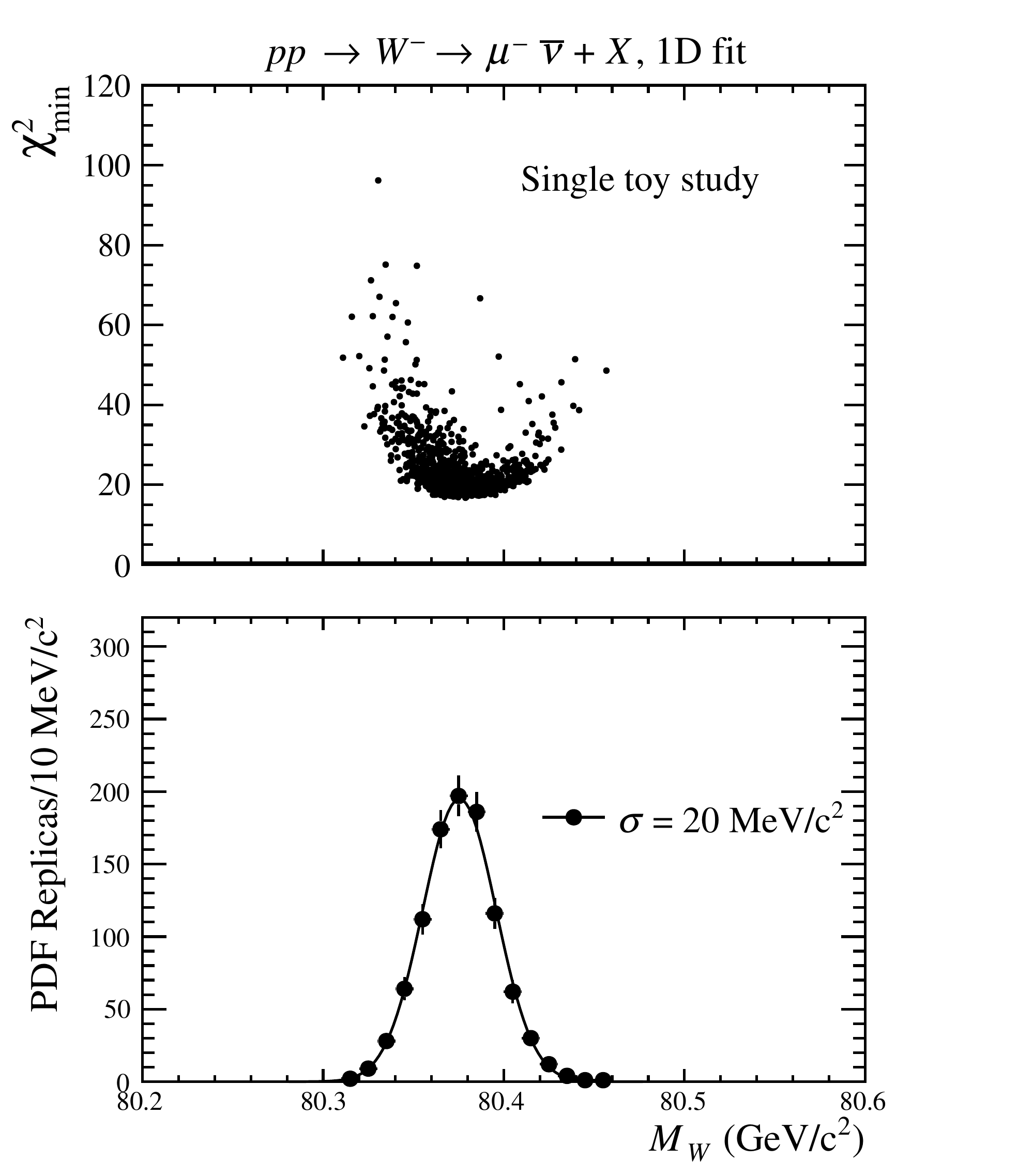}}
\caption{Upper: the distribution of the $\chi^2$ versus $M_W$ for a fit to a single toy dataset, which assumes the LHCb Run 2 statistics, with each of the 1000 NNPDF3.1 replicas. Lower: the distribution of the $M_W$ values with a Gaussian fit function overlaid.}\label{fig:corr_chi2minMw_1D}
\end{figure}

Fig.~\ref{fig:corr_chi2minMw_1D} shows, separately for the two $W$ charges, how the results of a fit to a single toy dataset vary with the PDF replica used in the templates. 
Forty bins in $p_T^\mu$ (with bin width of 0.5\,GeV/c) are used in the template fit. 
The fitted $M_W$ values follow approximately Gaussian distributions with widths of 15 (20)\,MeV/c$^2$ for the $W^+$ ($W^-$).
The broadly parabolic distributions of the best-fit $\chi^2$ ($\chi^2_{\text{min}}$) versus $M_W$ indicate that the PDF replicas that most severely bias $M_W$ tend to give a measurably poorer fit quality.
Before evaluating how this information could be used to constrain the PDF uncertainty let us first try to understand in more detail the underlying mechanism behind the PDF uncertainty.
\section{Understanding the PDF uncertainties}
\label{sec:pdf_uncertainties}
\begin{figure}
\centerline{
\includegraphics[width = 0.52\textwidth]{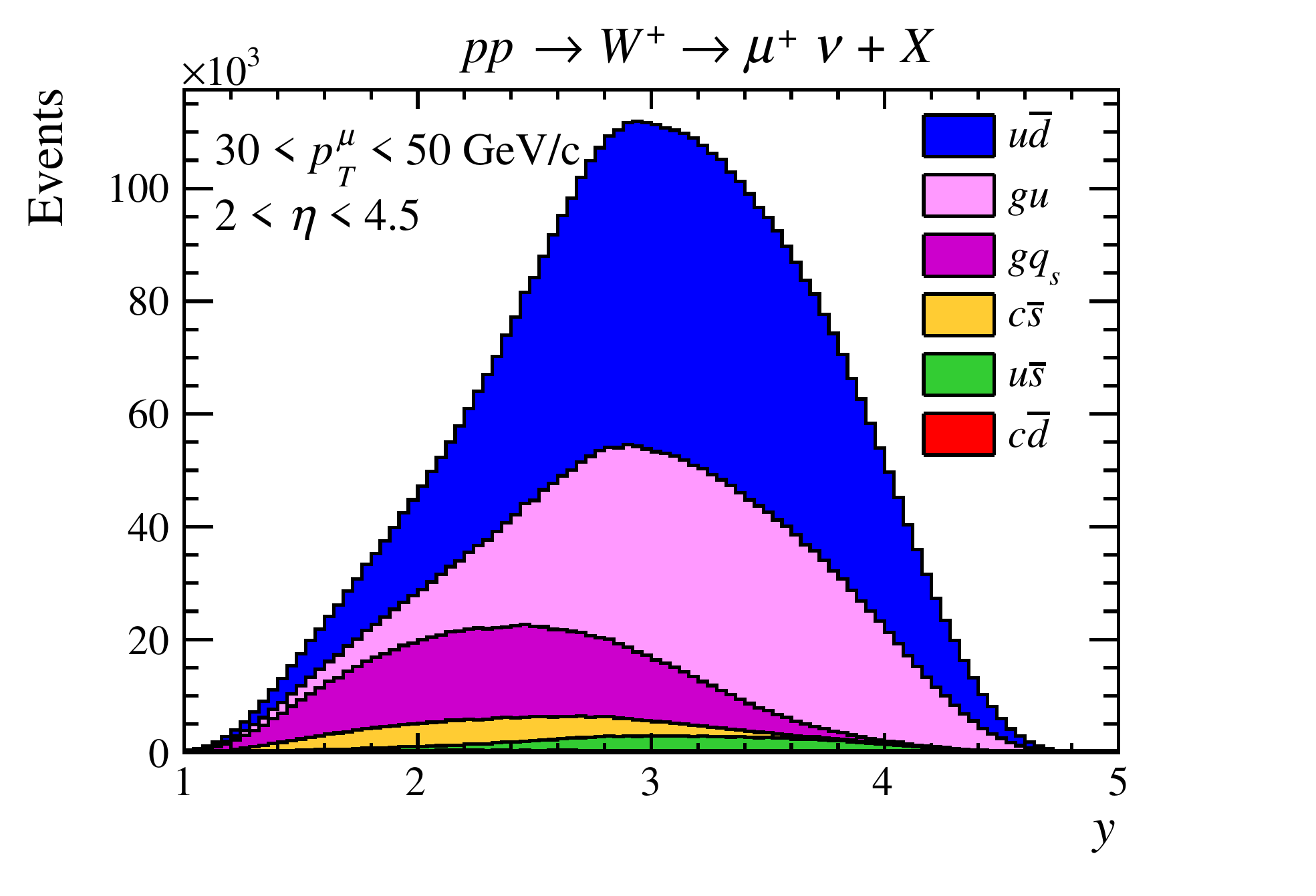}\hspace{-2mm}\includegraphics[width = 0.52\textwidth]{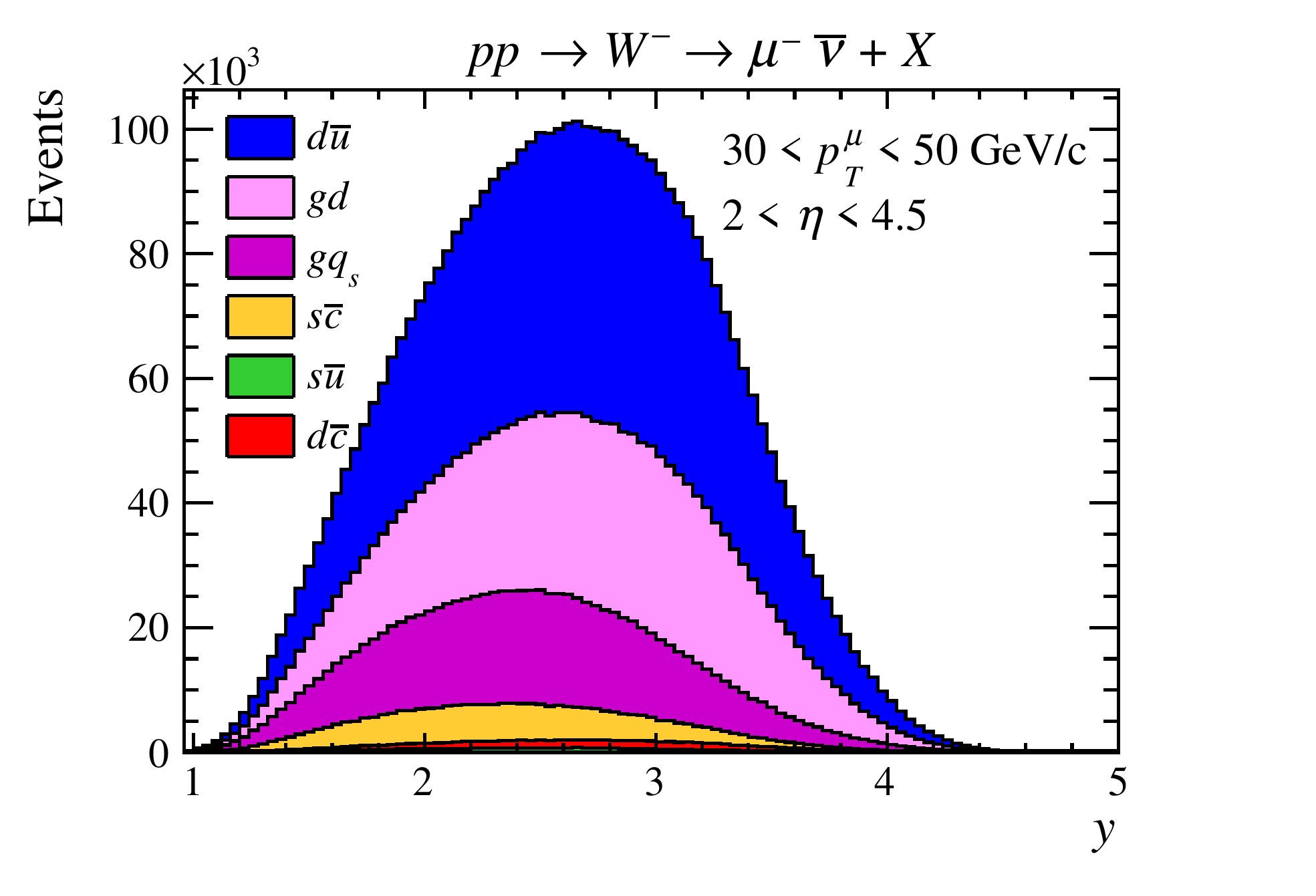}}
\caption{The (left) $W^+$ and (right) $W^-$ rapidity distributions decomposed into the main partonic subprocesses.}\label{fig:THStack_Wy}
\end{figure}

Fig.~\ref{fig:THStack_Wy} shows how the different partonic subprocesses contribute to the cross-section for $W$ production as a function of rapidity ($y$). 
The dominant $W^+$($W^-$) production subprocesses involve valence $u$($d$) quarks. Annihilation of gluons with sea quarks ($gq_s$) contributes for around a 20\% factor. Contributions from only second generation quarks annihilation are below 10\% or so.
%Fig.~\ref{fig:THStack} shows the parton flavour composition as a function of Bjorken $x$.
%There is a clear dominance of valence $u$($d$) quarks as the higher $x$ parton in $W^+$($W^-$) production. 
%The lower $x$ parton is usually a $\bar{d}$($\bar{u}$) sea quark in $W^+$($W^-$) production,
%but there is also a sizeable contribution from gluons, particularly in the $W^-$ case.

\begin{figure}
\centering
\includegraphics[width=0.5\textwidth]{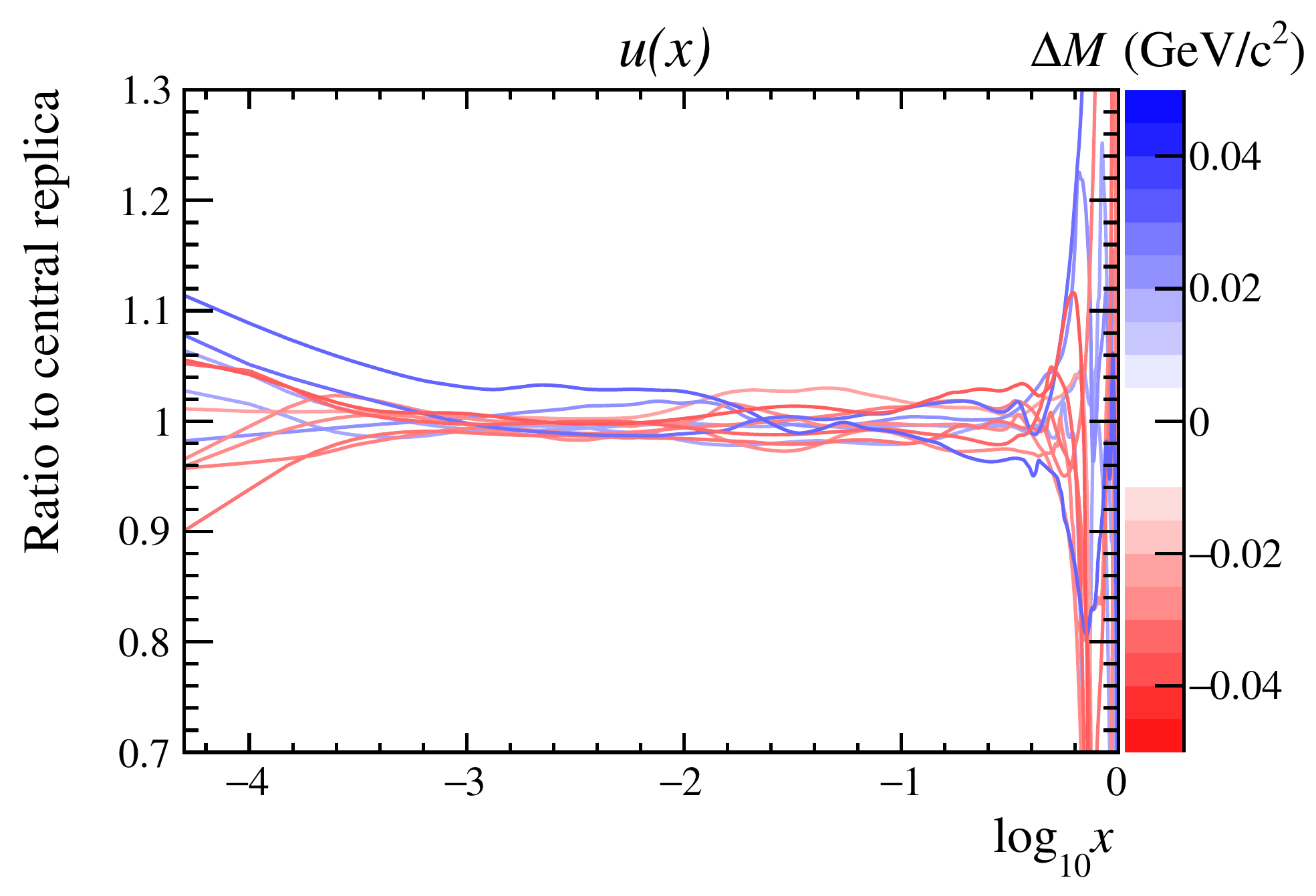}\includegraphics[width=0.5\textwidth]{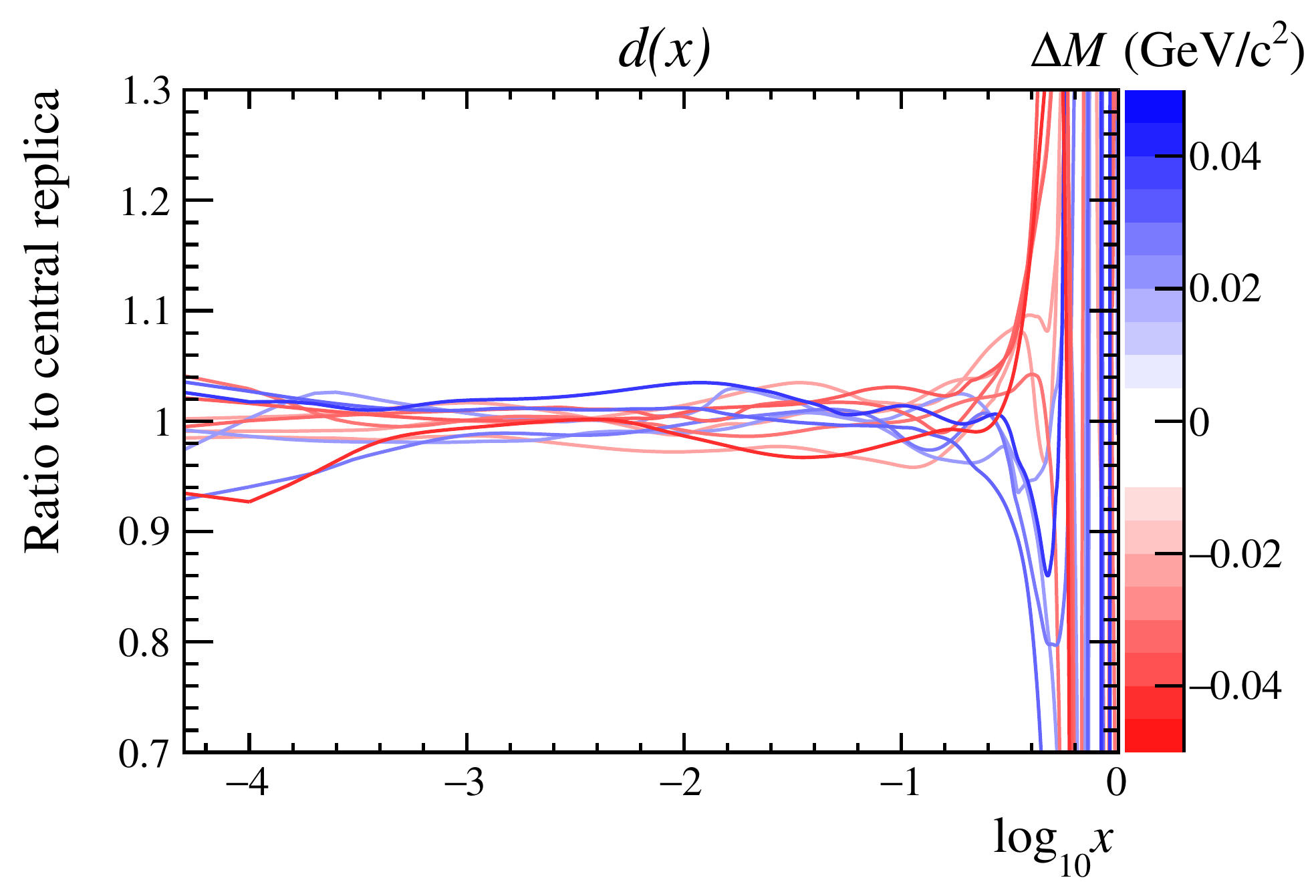}
\includegraphics[width=0.5\textwidth]{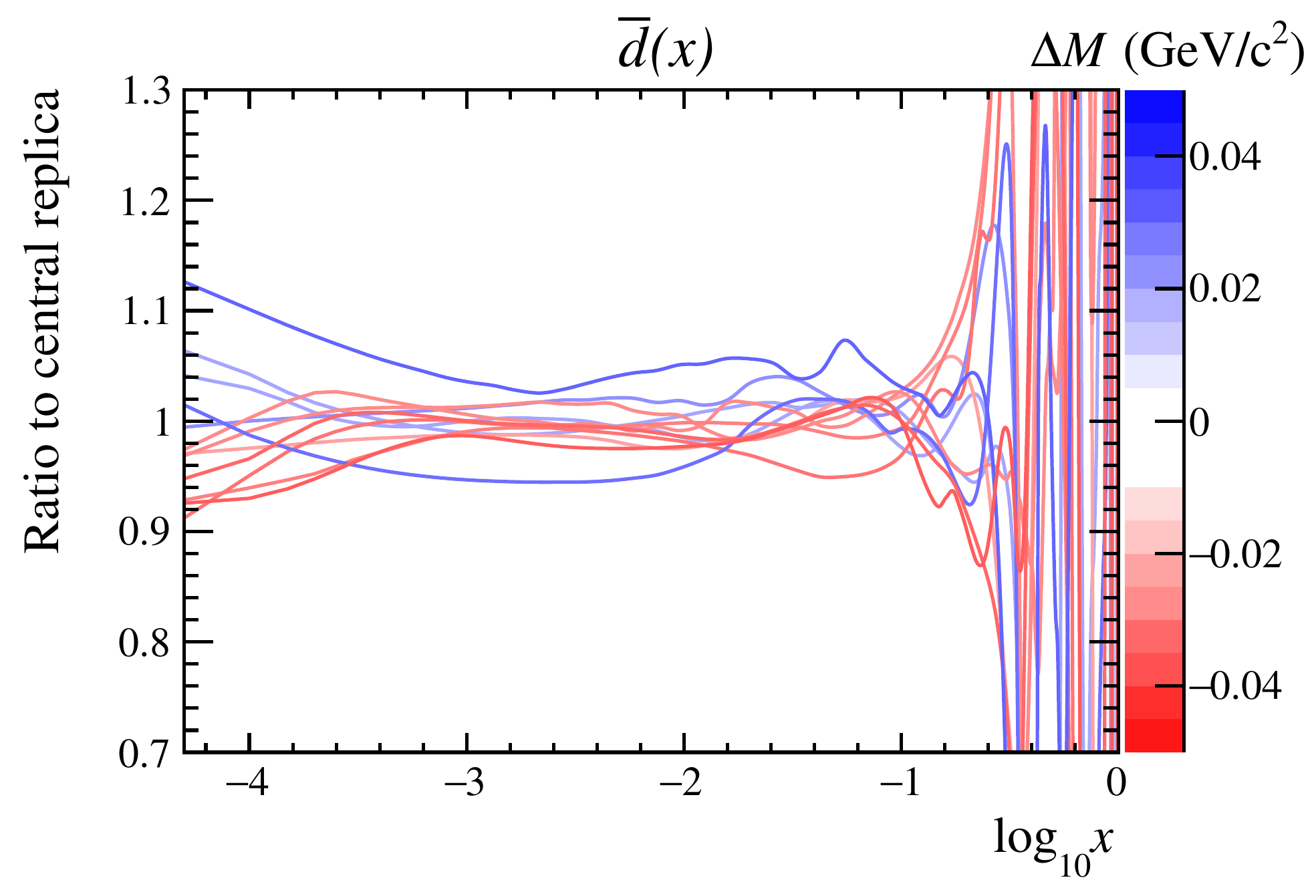}\includegraphics[width=0.5\textwidth]{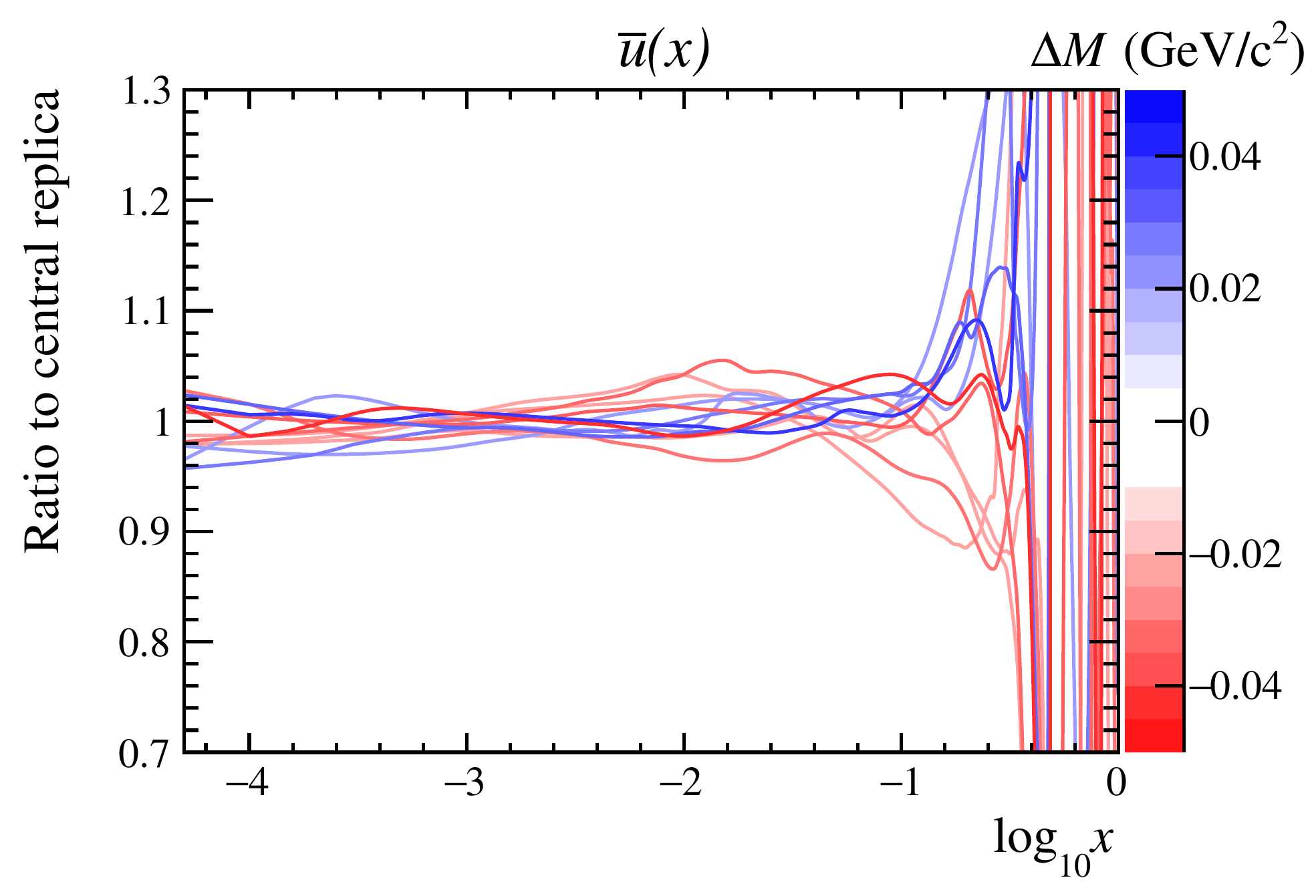}
\caption{The ratios of a subset of NNPDF3.1 replicas with respect to the central replica, for the $x$ dependence of the (clockwise from upper left) $u$, $d$, $\bar{u}$ and $\bar{d}$ PDFs. Each line is marked with a colour indicating the shift of the $M_W$ value
determined from a fit to the $p_T^\mu$ distribution of a single toy dataset. For clarity, the replicas for which the shift in $M_W$ is close to zero are not drawn.}\label{fig:PDFbands}
\end{figure}

Since the $u$, $\bar{d}$, $d$ and $\bar{u}$ species seem to be the most important it is interesting to see 
if there are any obvious patterns in their respective PDFs for the replicas corresponding to biased $M_W$ determinations. 
The final results are derived using the full set of 1000 NNPDF3.1 equiprobable replicas but, for visual purposes, the studies in this section make use of a subset of them. Fig.~\ref{fig:PDFbands} shows how the $x$ dependencies of the 
$u$, $\bar{d}$, $d$ and $\bar{u}$ PDFs vary between the subset of replicas.
Each line is a ratio with respect to the central replica, and is assigned a colour according to the bias in $M_W$ as evaluated using the method described in Sect.~\ref{sec:toy_method}. For clarity, the replicas for which the shift in $M_W$ is close to zero ($|\Delta M|< 10$ MeV/c$^2$) are not drawn. In the study of the single partonic species, only the relevant $W$ charges templates are included in the fit. 
No obvious patterns can be seen in the $u$ and $\bar{d}$ PDFs, which dominate $W^+$ production.
However, a clear pattern can be seen for the high-$x$ (above $x \sim 0.1$) $d$ PDF,
whereby the replicas that tend to bias $M_W$ upwards (downwards) tend to have a smaller (larger) parton density.
A qualitatively similar pattern, though with the opposite sign, is seen in the $\bar{u}$ PDF.
%\begin{figure}
%\centering
%\includegraphics[width = 0.5\textwidth]{figures_tempPDF/THStack_Wp}\includegraphics[width = 0.5\textwidth]{figures_tempPDF/THStack_Wm}
%\caption{The Bjorken $x$ distributions of the partons involved in the (left) $W^+$ and (right) ($W^-$) hard processes.
%The main parton species are stacked upon each other.}\label{fig:THStack}
%\end{figure}

\begin{figure}
\centerline{\hspace{1.5mm}\includegraphics[width = 0.5\textwidth]{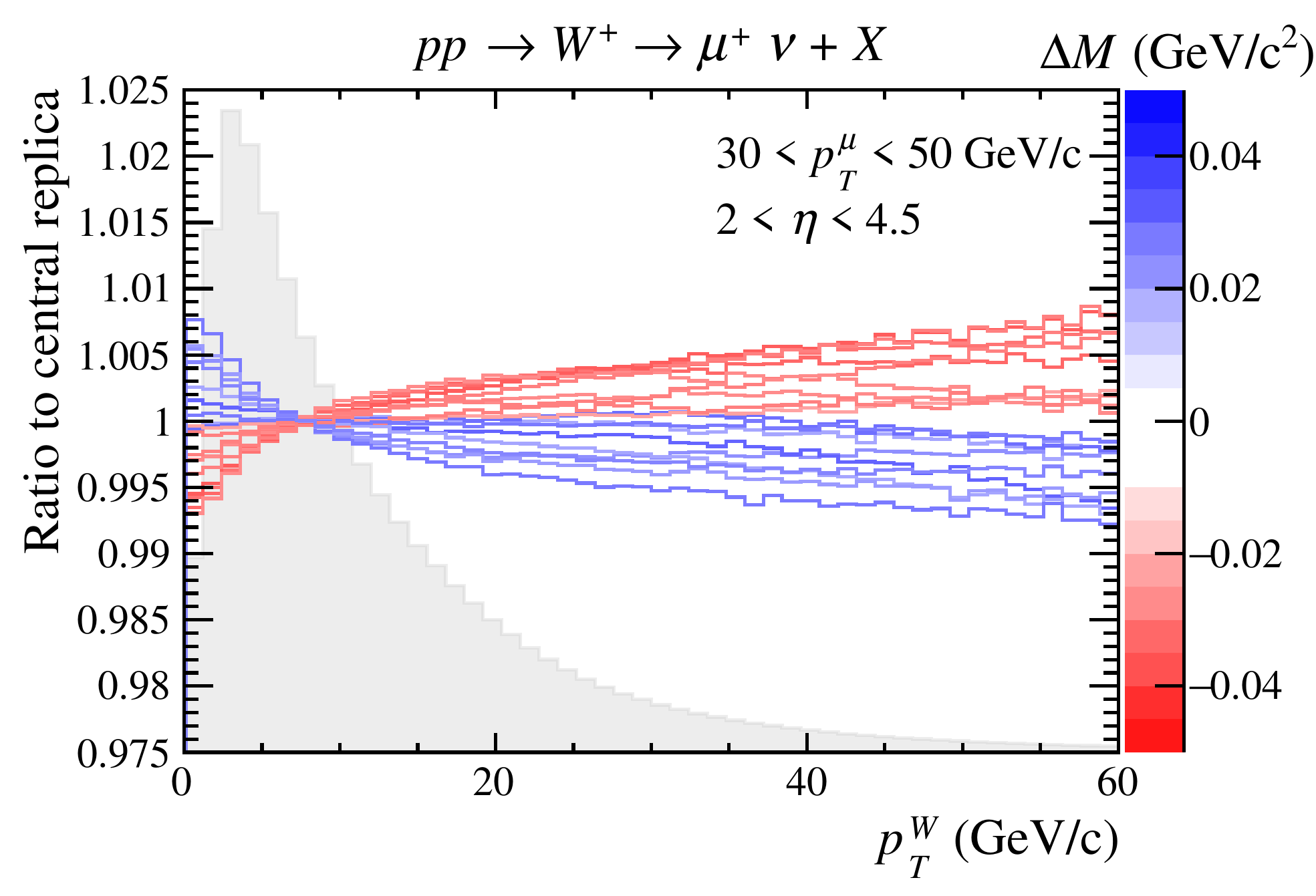}\hspace{-1.7mm} \includegraphics[width = 0.5\textwidth]{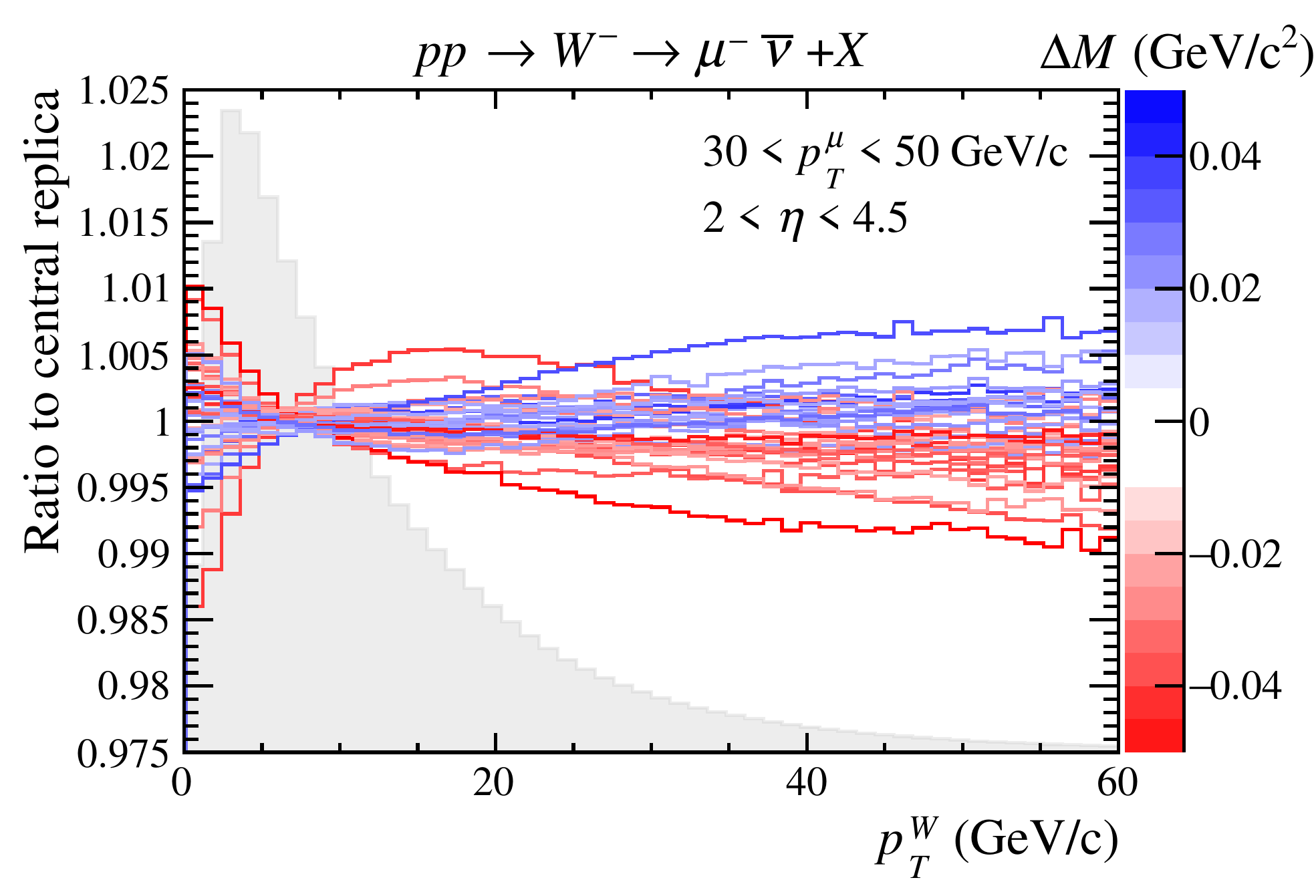}}
\centerline{
\includegraphics[width = 0.5\textwidth]{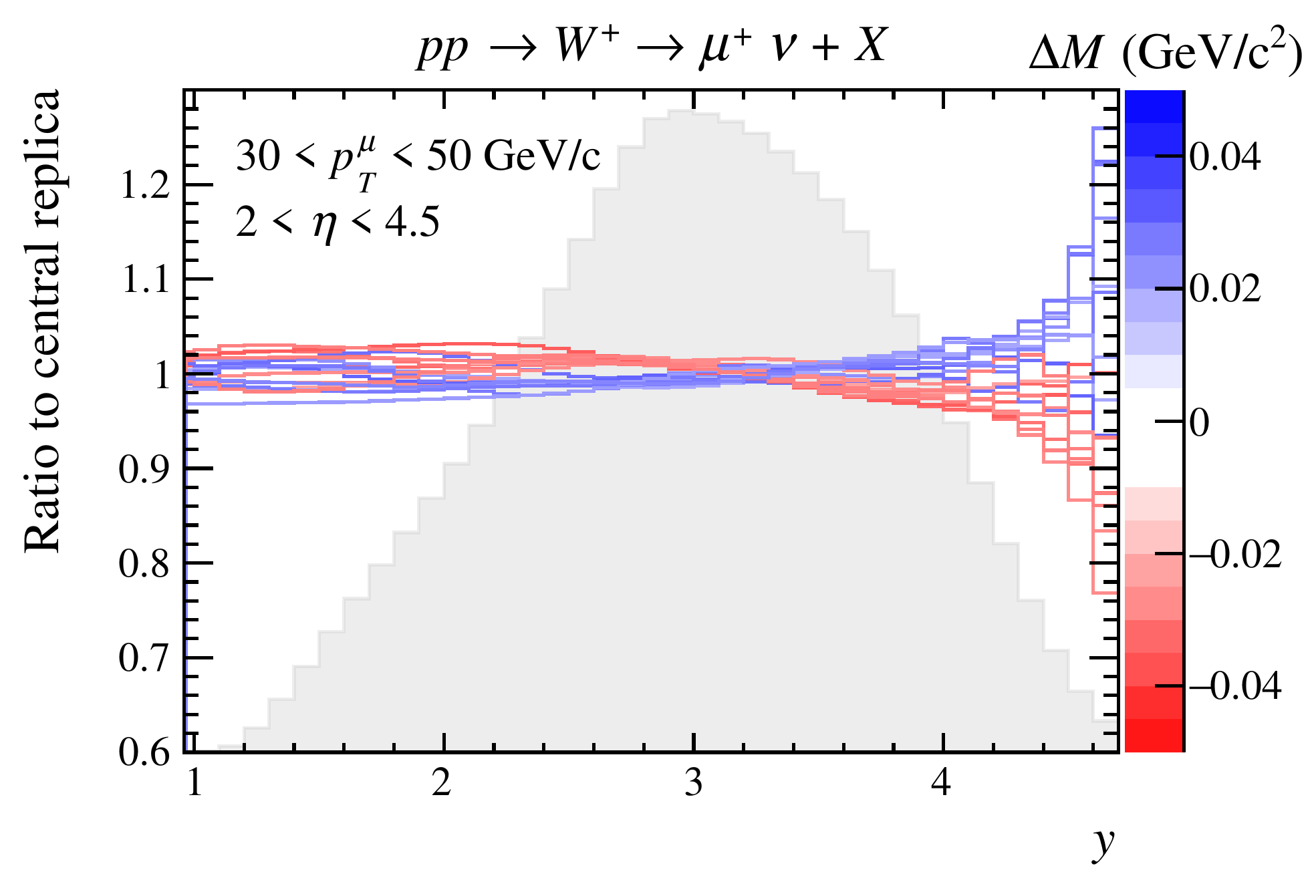}\includegraphics[width = 0.5\textwidth]{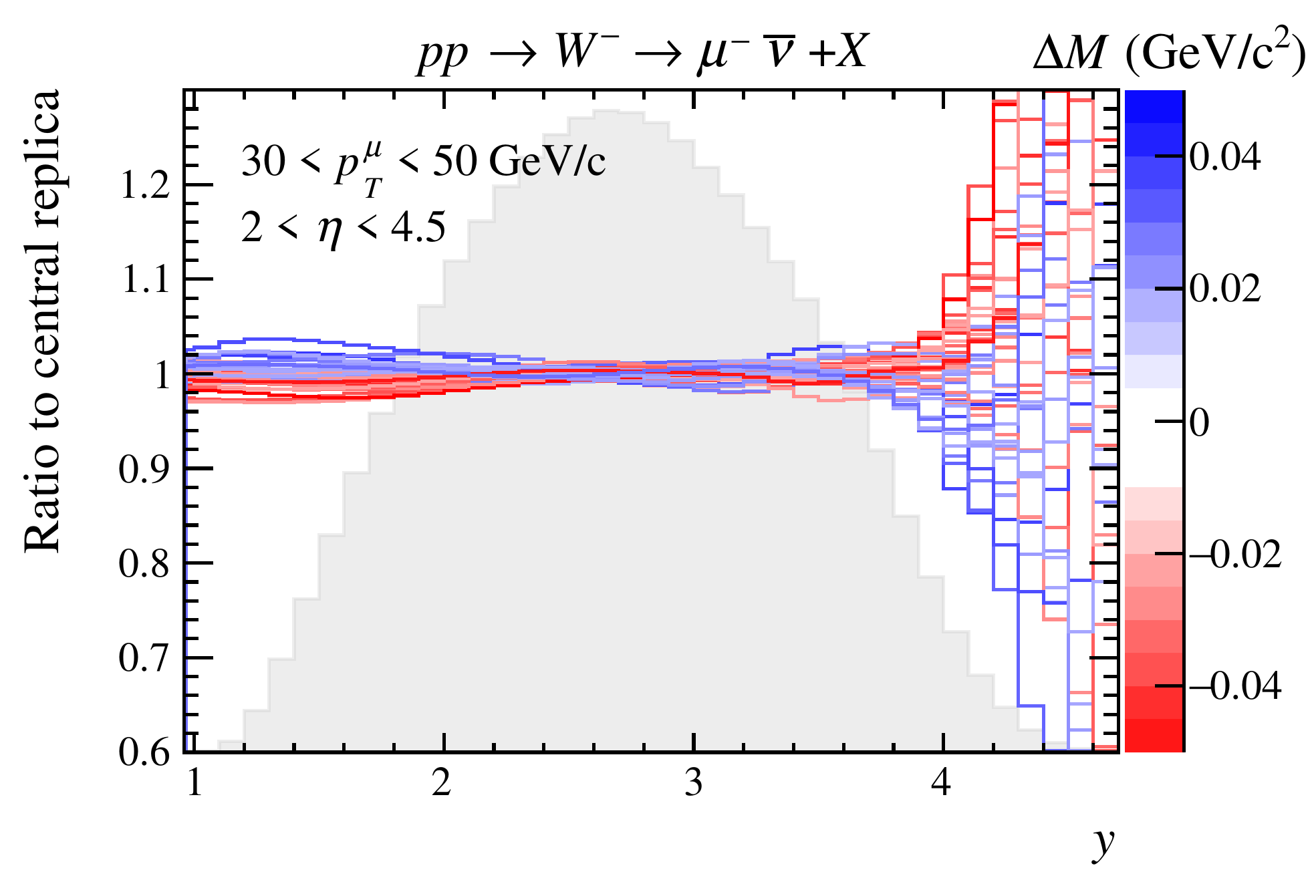}}
\centerline{
 \includegraphics[width = 0.5\textwidth]{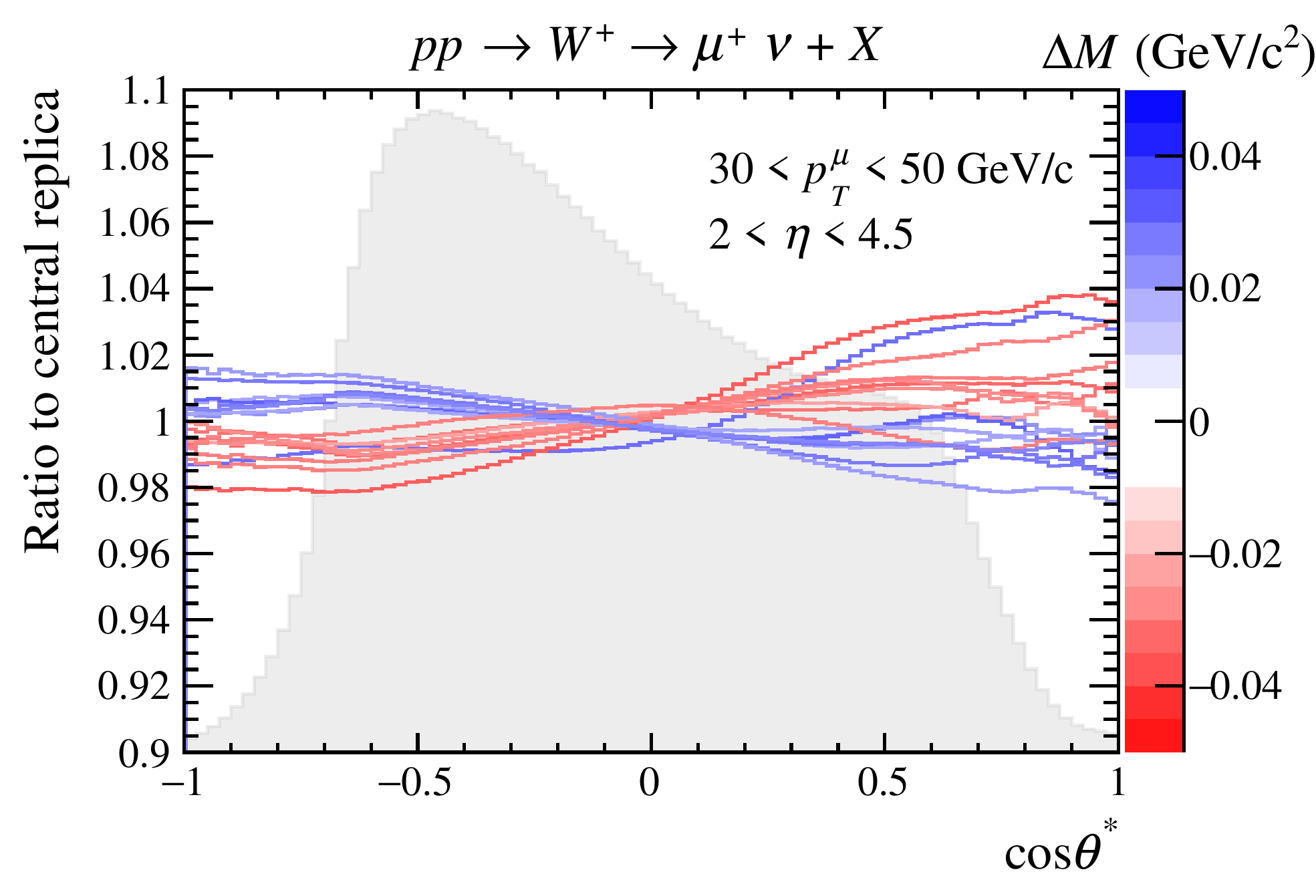}\includegraphics[width = 0.5\textwidth]{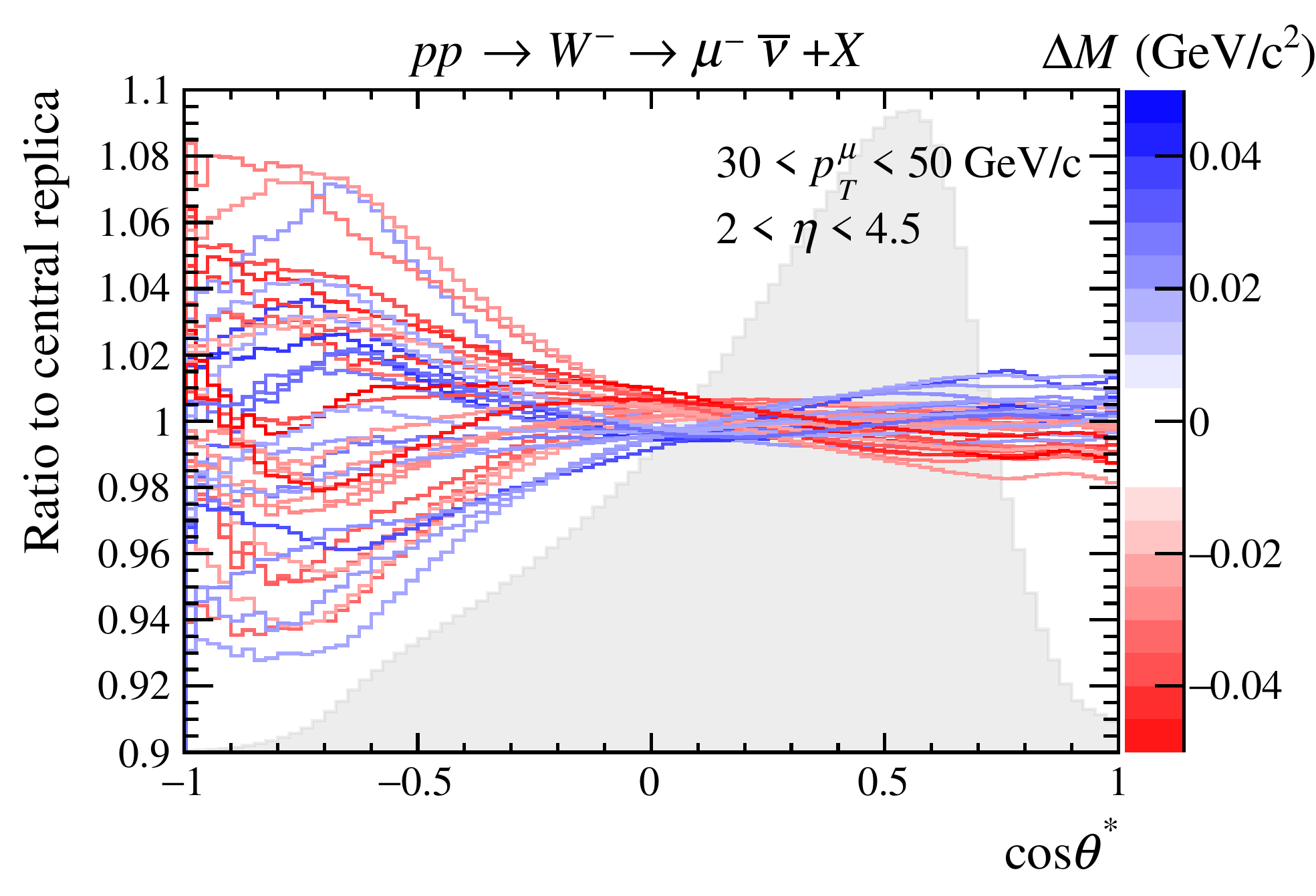}}
\caption{The variations in the shapes of the $p_T^W$, $y$ and $\cos\theta^*$ distributions predicted with a subset of  NNPDF3.1 replicas.
Each line is marked with a colour indicating the shift of the $M_W$ value determined from a fit to the $p_T^\mu$ distribution of a single toy dataset.  
For clarity, the replicas for which the shift in $M_W$ is close to zero are not drawn.}\label{fig:W_PDFband}
\end{figure}

\begin{figure}
\centerline{
\includegraphics[width = 0.5\textwidth]{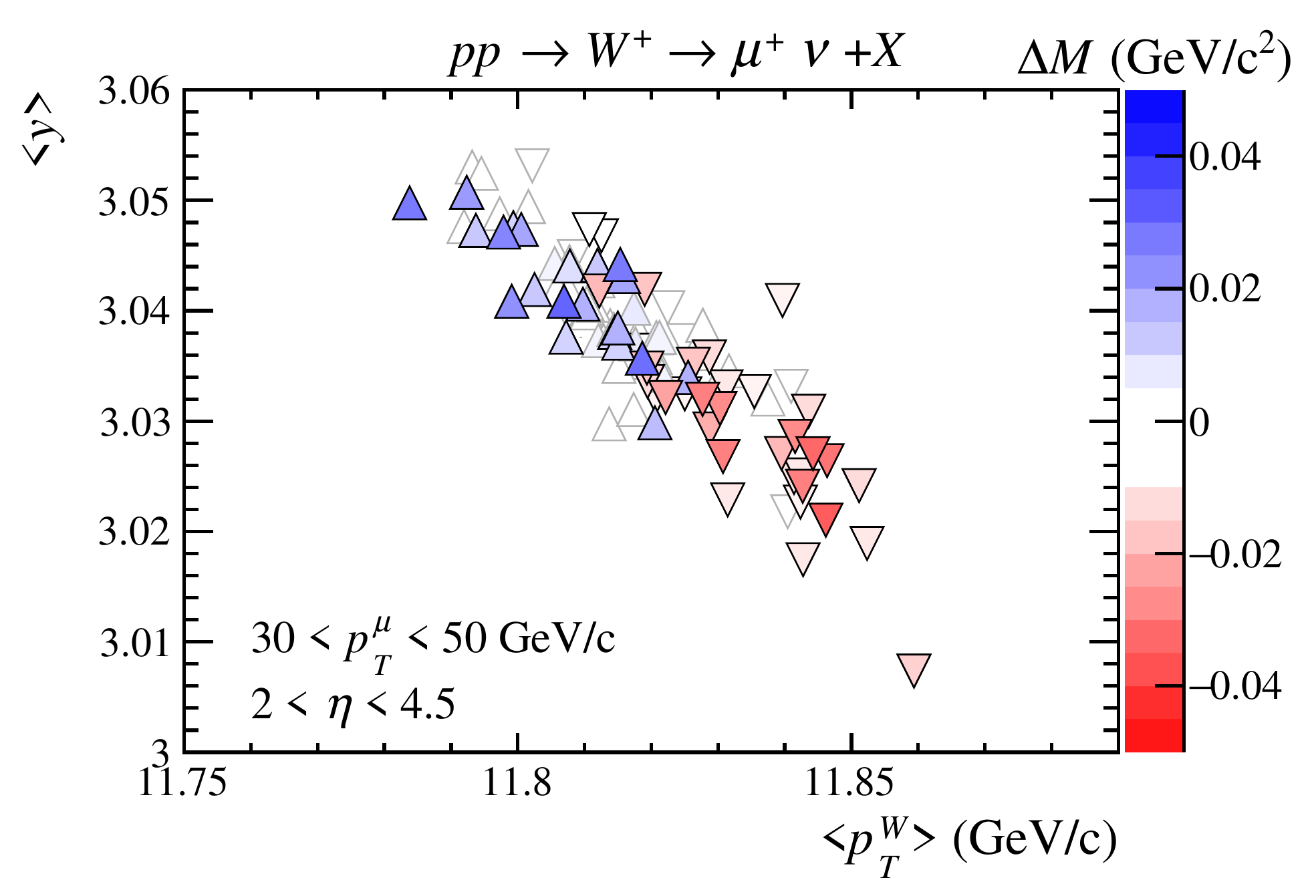} \includegraphics[width = 0.5\textwidth]{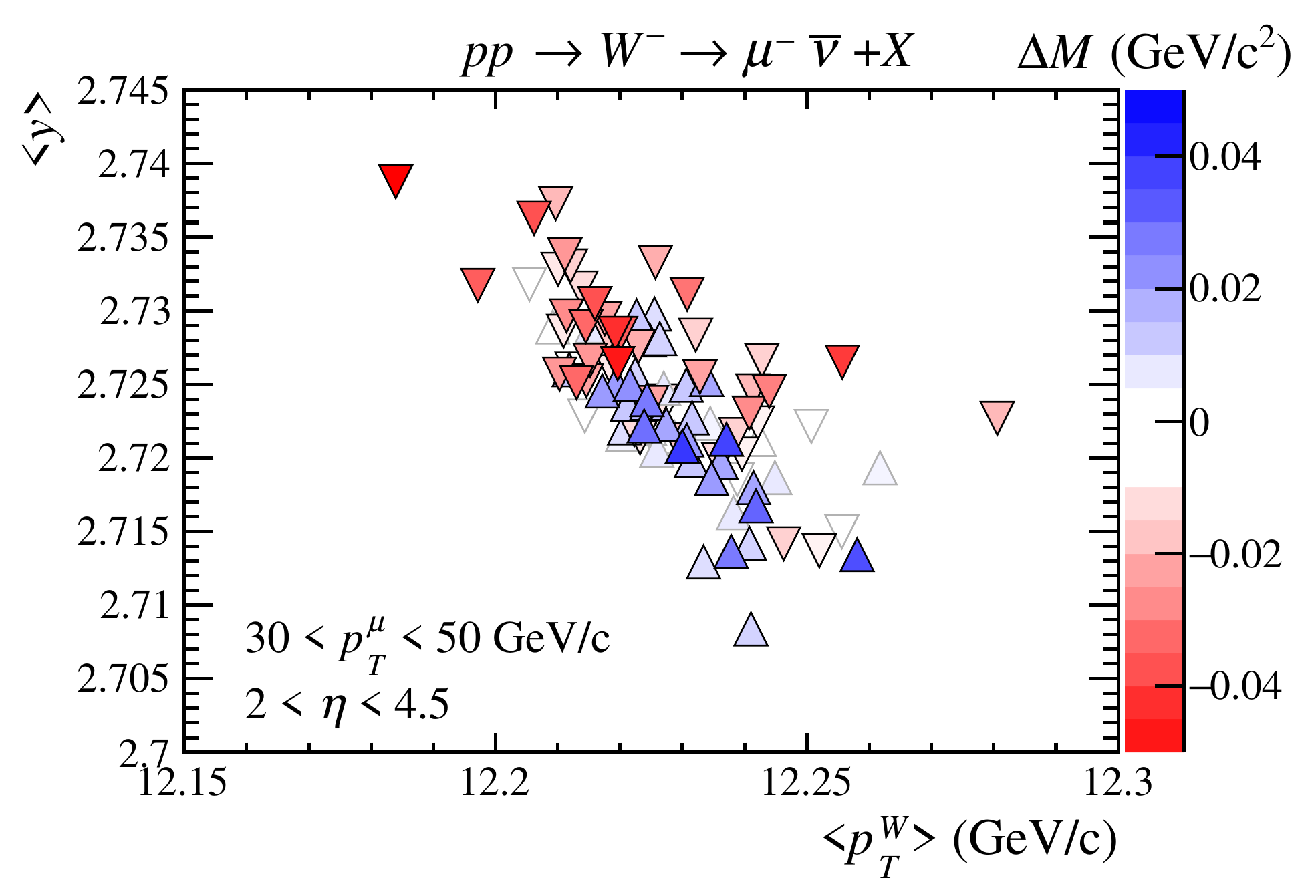}}
\caption{The distributions of $\langle p_T^W\rangle$ and $\langle y\rangle$ for a subset of replicas of the NNPDF3.1 set. Each marker is assigned a colour according to the shift of the $M_W$ value determined from a fit to the $p_T^\mu$ distribution of a single toy dataset. The markers drawn with an up(down) pointing triangle correspond to $\Delta M$ values greater(less) than zero.} \label{fig:corr_W_Wpm}
\end{figure}

The PDF uncertainty on the $M_W$ measurement arises because the $p_T^\mu$ distribution depends on the $W$ production kinematics, 
which are characterised by the transverse momentum ($p_T^W$), rapidity and polarisation.
As a proxy for the polarisation, the distribution of the angle $\theta^*$ in the Collins-Soper frame~\cite{CSF} can be considered.
Fig.~\ref{fig:W_PDFband} shows how the $p_T^W$, $y$ and $\cos\theta^{*}$ distributions vary between a subset of  NNPDF3.1 replicas.
Each line is assigned a colour according to the bias in $M_W$ for that replica.
The underlying shapes of the distributions are also indicated by the filled histograms.
A particularly striking pattern can be seen in the variation of the $y$ distributions.
The replicas that bias $M_W$ upwards (downwards) tend to enhance (suppress) the $W^+$ cross-section at large rapidities.
The opposite is seen for the $W^-$. 
Other clear patterns, though with smaller absolute variations, can be seen in the $p_T^W$ and $\cos\theta^{*}$ projections.
It is instructive to consider the two-dimensional projections of these patterns.
Fig.~\ref{fig:corr_W_Wpm} shows the mean of the $y$ distribution versus the mean of the $p_T^W$ distribution.
Each point represents a single NNPDF3.1 replica using the already described $M_W$ dependent colour scale.
There is a clear anticorrelation between the changes in the shapes of the $y$ and $p_T^W$ distributions which is expected from the kinematics and is enhanced by the forward acceptance cuts applied to the lepton, 
but further patterns can be seen in the colour distribution.
In the $W^+$ case, the replicas that bias $M_W$ upwards (downwards) tend to predict larger (smaller) $\langle y \rangle$ values and smaller (larger) $\langle p_T^W \rangle$ values.
The opposite pattern is seen for the $W^-$ case. 
These striking patterns are helpful in understanding how biases in $M_W$ are correlated to the underlying $W$ production kinematics. 

\begin{figure}
\centerline{
\includegraphics[width = 0.5\textwidth]{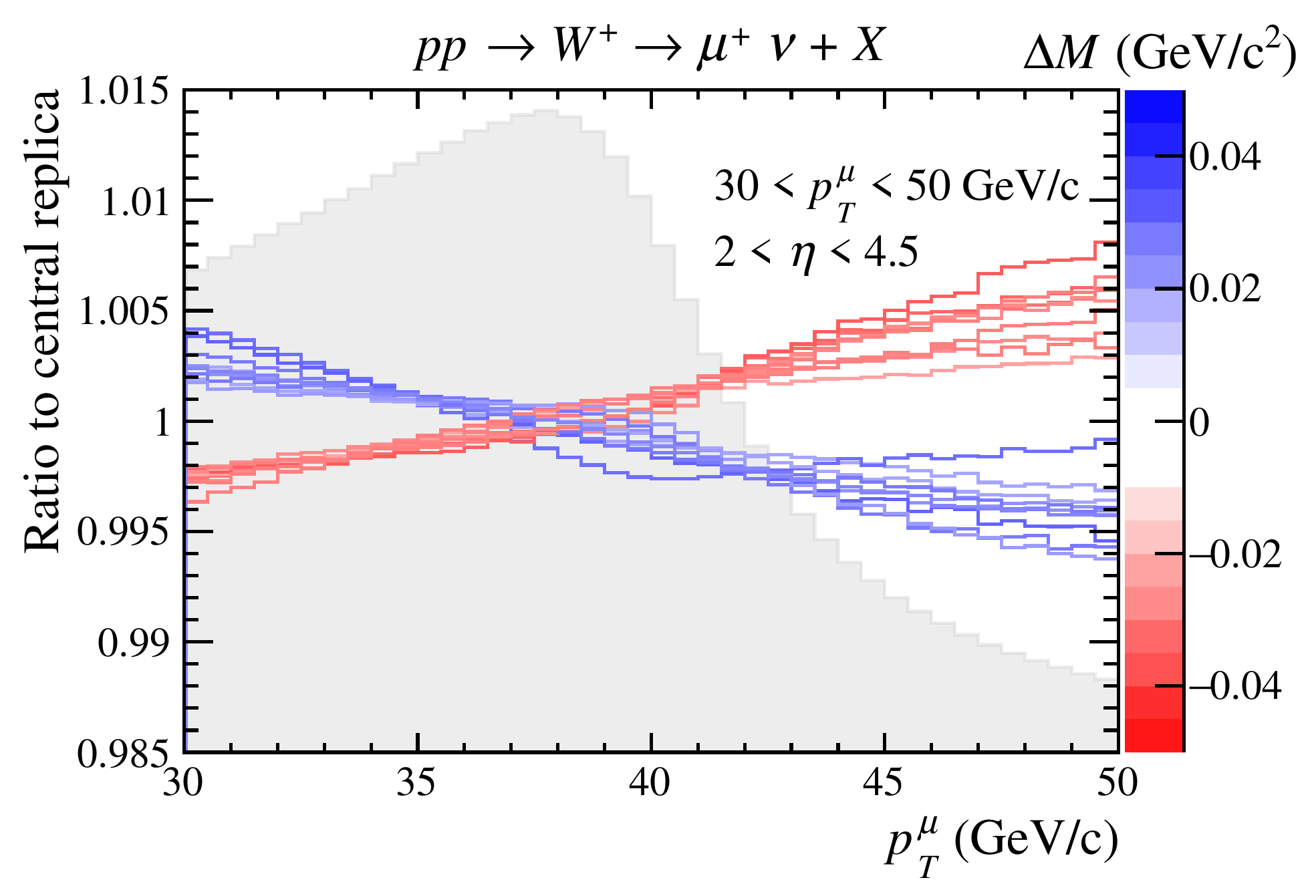}\includegraphics[width = 0.5\textwidth]{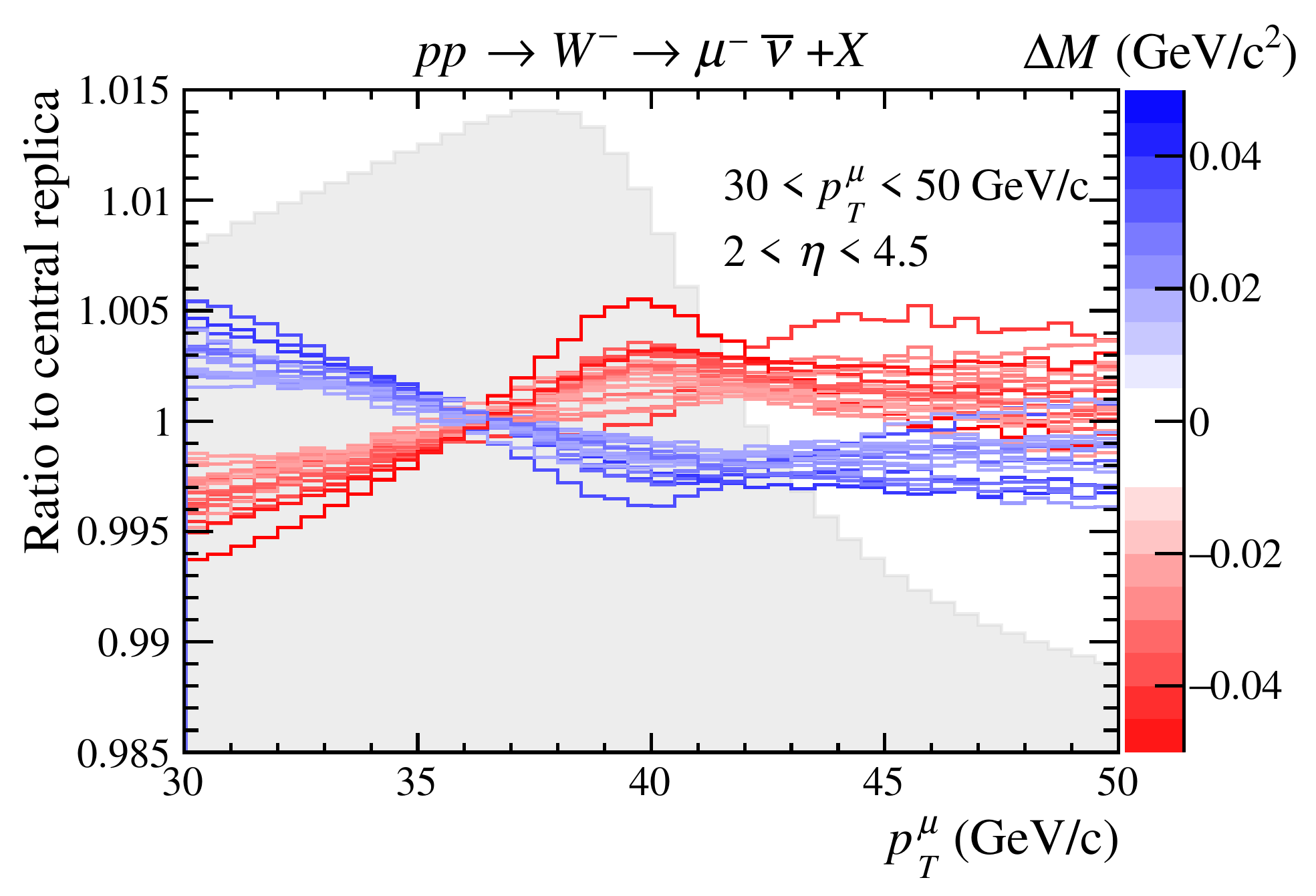}}
\centerline{
\includegraphics[width = 0.5\textwidth]{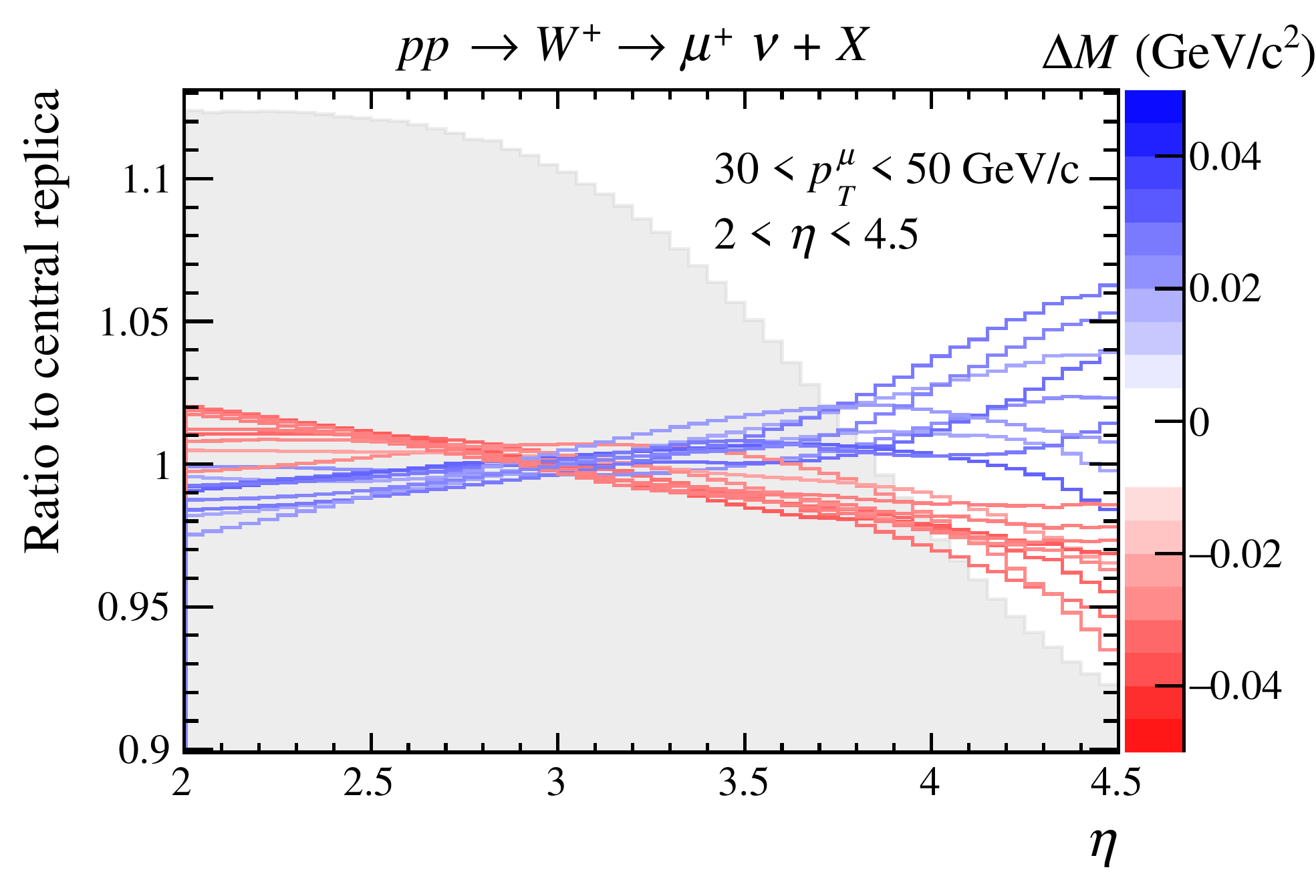}\includegraphics[width = 0.5\textwidth]{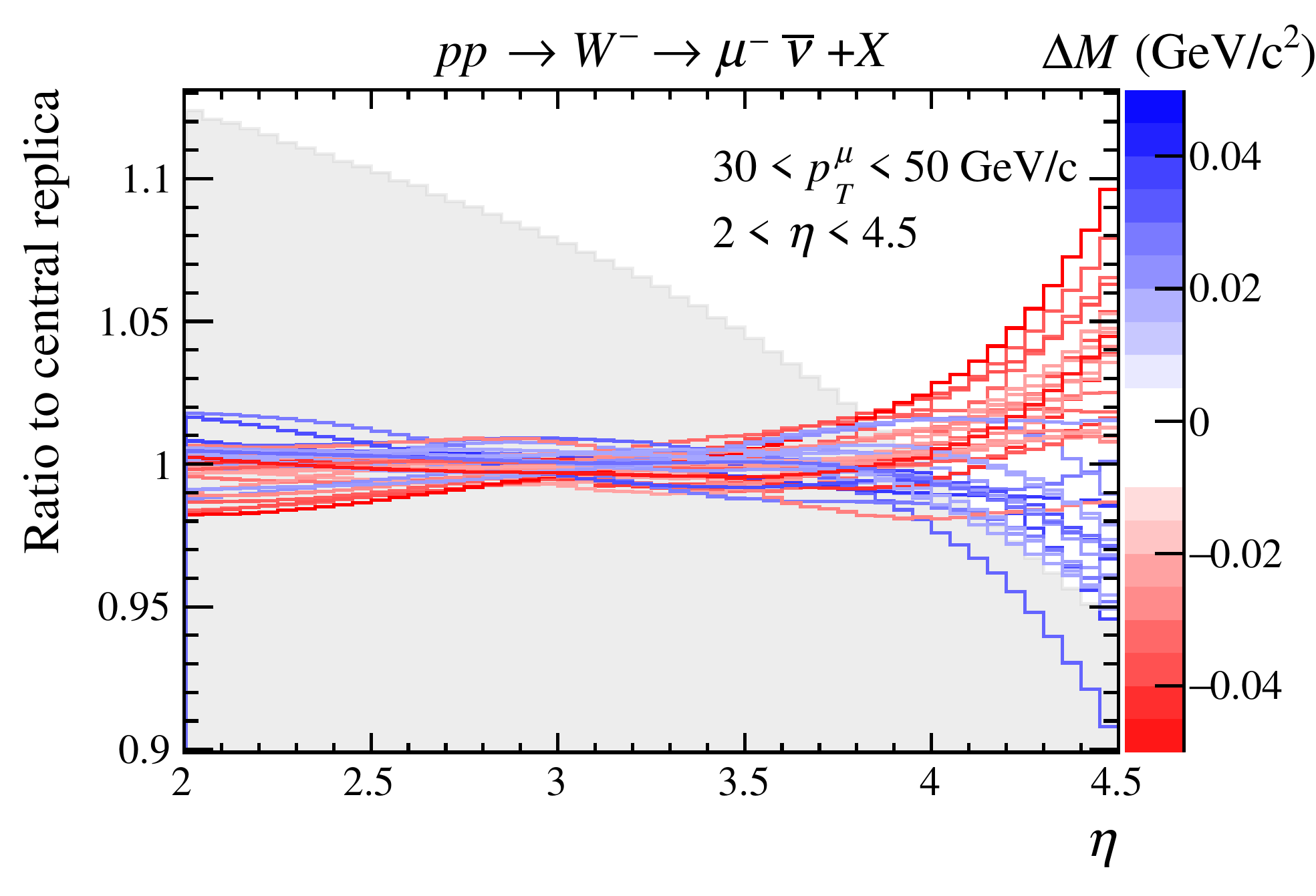}}
\caption{The variations in the shapes of the $p_T^\mu$ and $\eta$ distributions predicted with a subset of NNPDF3.1 replicas.
Each line is  marked with a colour indicating the shift of the $M_W$ value determined from a fit to the $p_T^\mu$ distribution of a single toy dataset. For clarity, the replicas for which the shift in $M_W$ is close to zero are not drawn.}\label{fig:mu_PDFband}
\end{figure}

\begin{figure}[h]
\centerline{
\includegraphics[width = 0.5\textwidth]{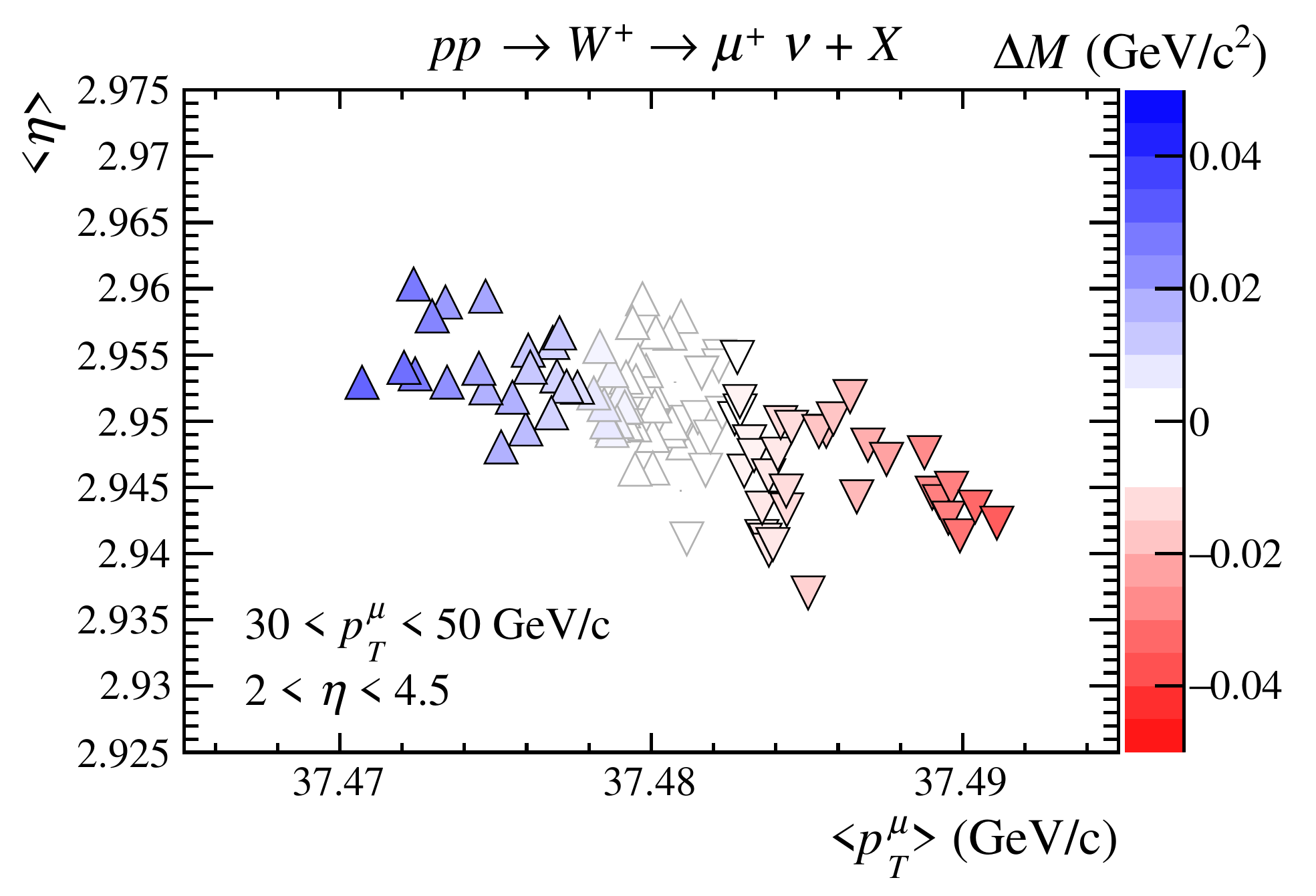}\includegraphics[width = 0.5\textwidth]{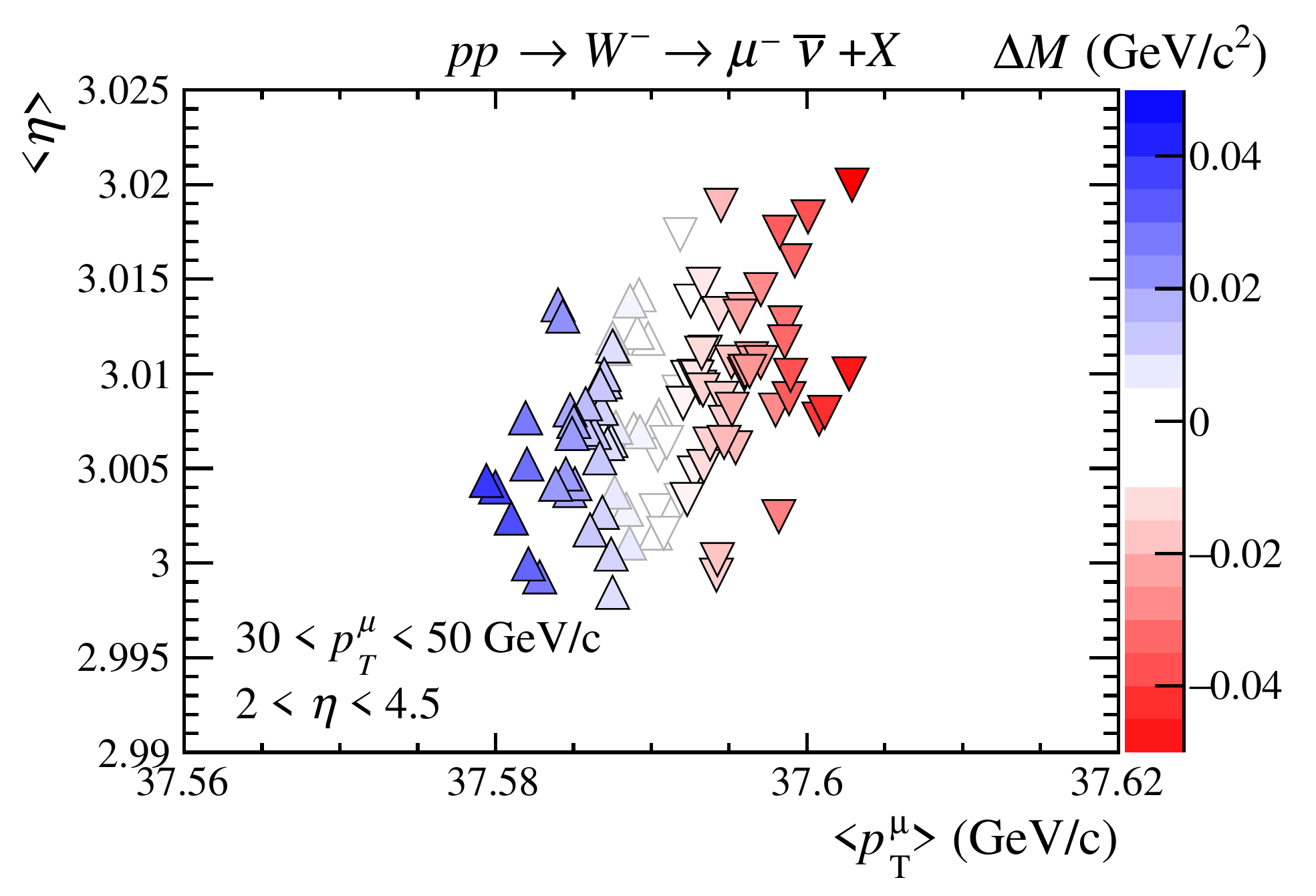}}
\caption{The distribution of $\langle p_T^\mu\rangle$ and $\langle\eta\rangle$ produced using a subset of replicas of the NNPDF3.1 set and divided by the central replica. Each marker is assigned a colour according to the shift of the $M_W$ value determined from a fit to the $p_T^\mu$ distribution of a single toy dataset. The markers drawn with an up(down) pointing triangle correspond to $\Delta M$ values greater(less) than zero.}\label{fig:mu_PDFcorr}
\end{figure}

Our attention is now switched to the muon kinematic distributions.
%encouraged by interesting studies on disentangling the $W$ kinematics from the measurable muon observables in the ATLAS/CMS acceptance~\cite{Manca}.
Ref.~\cite{Manca} showed that correlated changes in the shapes of the $\eta$ and $p_T^\mu$ distributions in the phase-space acceptance of ATLAS and CMS can be used to further constrain the PDFs. It is therefore interesting to consider a similar approach in the LHCb phase-space acceptance.
Fig.~\ref{fig:mu_PDFband} shows how the muon $p_T^\mu$ and $\eta$ distributions vary with the PDF replicas.
As expected the replicas that bias $M_W$ upwards (downwards) correspond to a decrease (increase) in the predicted cross-section at high $p_T^\mu$ with respect to low $p_T^\mu$.
%An intriguing observation, however, is that the replicas that provide the largest bias on $M_W$ cause a correlated change in the shape of the $\eta$ distribution. 
An intriguing observation, however, is that the replicas that provide the largest bias on $M_W$ change not only the shape of the $p_T^\mu$ distribution but also that of the $\eta$ distribution.
This is a {\em measurable} change of up to several percent, which could be exploited to constrain the PDF uncertainty.
Fig.~\ref{fig:mu_PDFcorr} shows the mean $p_T^\mu$ versus the mean $\eta$ for each replica, with the $M_W$ dependent colour scale as before. 
The replicas that bias $M_W$ tend to be clearly separated in this two-dimensional plane, which encourages us to consider exploiting this information
to constrain the PDF uncertainty.
\FloatBarrier

\section{PDF uncertainty reduction}
\label{sec:sigmapdf_reduction}
\begin{figure}
\centerline{   
\includegraphics[width=0.5\textwidth]{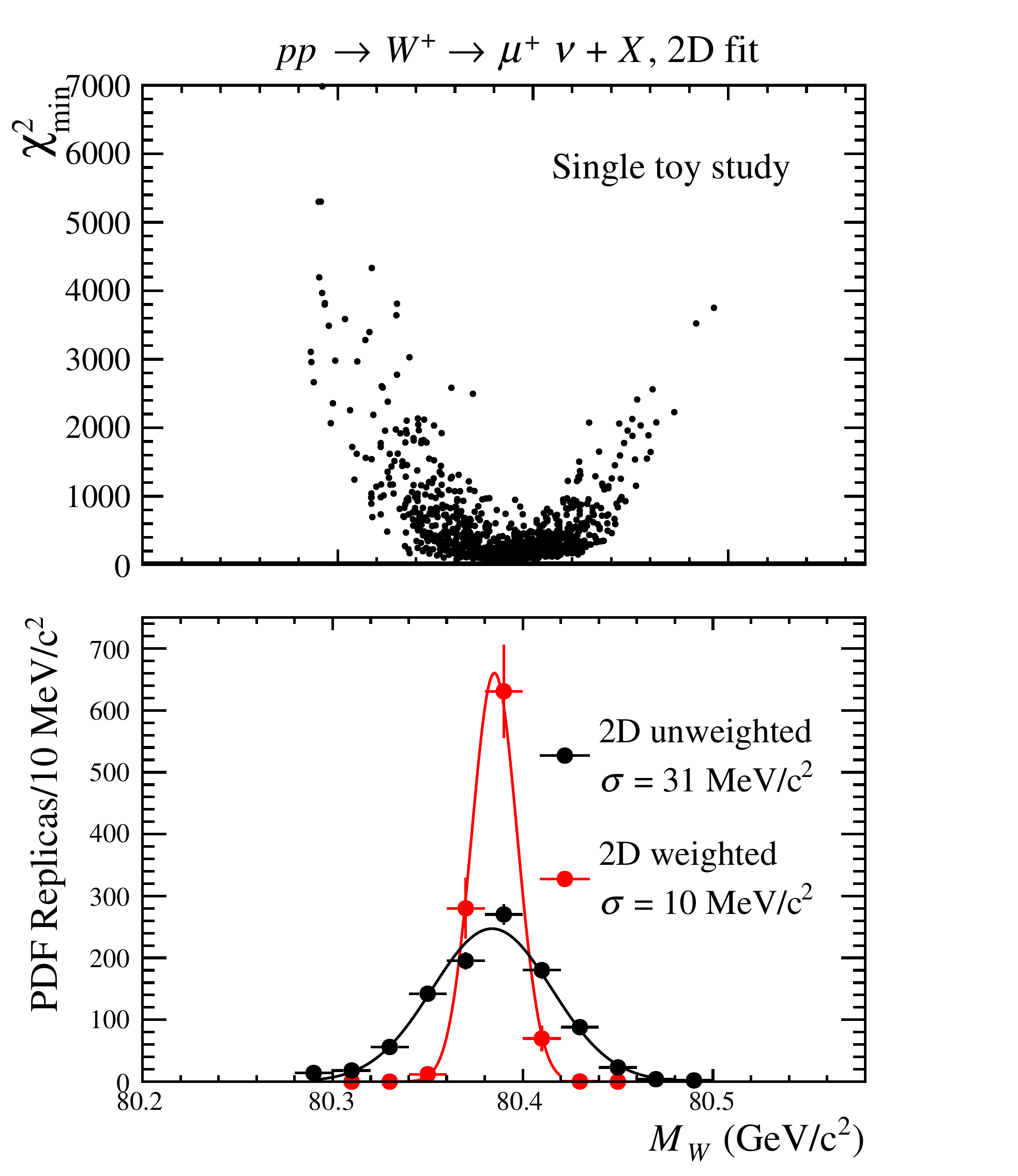}\includegraphics[width=0.5\textwidth]{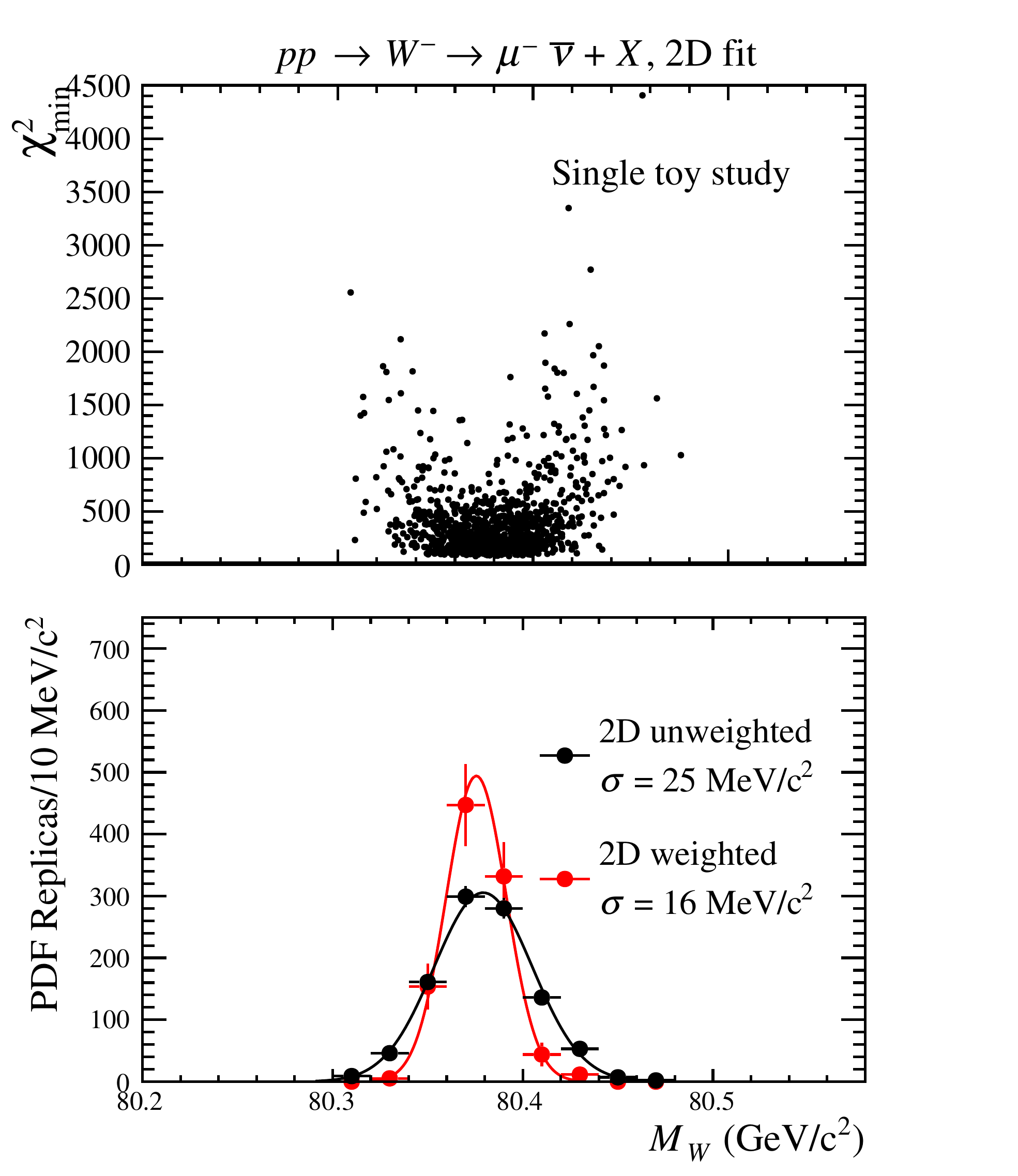}}
\caption{Upper: the distribution of the $\chi^2$ versus $M_W$ for a two-dimensional fit to a single toy dataset, which assumes the LHCb Run 2 statistics, with each of the 1000 NNPDF3.1 replicas. Lower: the distribution of the extracted $M_W$ values, with a Gaussian fit function overlaid, without (black) and with (red) weighting.}\label{fig:corr_chi2minMw_2D}
\end{figure}

In Sect.~\ref{sec:toy_method} it was noted that the traditional one-dimensional fit to the $p_T^\mu$ distribution already suggests a potential for \emph{in situ} constraints of the PDF uncertainty~\cite{Giele}.
%However, Sect.~\ref{sec:pdf_uncertainties} encourages the consideration of a fit to the two-dimensional ($p_T$ versus $\eta$) distribution. 
%One-dimensional and two-dimensional fits are now compared with and without the inclusion of replica weights.
The fit is now compared with and without the inclusion of replica weights.
Using the NNPDF prescription~\cite{NNPDFrew1,NNPDFrew2}, each replica is assigned a weight according to the best-fit $\chi^2$ ($\chi^2_{\text{min}}$) for a fit with $n$ degrees of freedom ($n$):
\begin{equation}
P(\chi^2_{\text{min}}) \propto \chi^{2^{\,(n-1)}}_{\text{min}}e^{-\frac{1}{2}\chi^2_{\text{min}}}. \label{eq:weight_shape}
\end{equation} 
This has the effect of disregarding replicas that are incompatible with the data. An alternative approach is to use the PDFs represented by Hessian eigenvectors and profile them in the analysis~\cite{hessian}. 
Sect.~\ref{sec:pdf_uncertainties} encourages the consideration of a fit to the two-dimensional ($p_T^\mu$ versus $\eta$) distribution to further constrain the PDF uncertainty. 
The two-dimensional fit uses three bins in $\eta$ within the ($2 < \eta < 4.5$) range and forty bins in $p_T^\mu$ within the ($30 < p_T^\mu <50$\,GeV/c) range already described.
Fig.~\ref{fig:corr_chi2minMw_2D} shows, separately for the $W^+$ and $W^-$ cases, the distribution of $M_W$ and $\chi^2$ values for the two-dimensional fit to a single toy dataset. 
The distributions of $M_W$ values are shown with and without the replica weights. In the $W^+$ case the width of weighted distribution is roughly a factor of three smaller than the unweighted distribution.
For the $W^-$ the width is reduced by roughly 50\%. The effective number of replicas after reweighting 
\begin{equation}
N_{\text{eff}} = \frac{(\Sigma_{i=1}^N w_i)^2}{\Sigma_{i=1}^N w^2_i} \quad \text{with} \quad w_i = P(\chi^2_{i, \text{min}}),
\end{equation}\label{eq:Neff}
where $N$ is the total number of replicas, gives an indication of the statistical reliability of the method. It is estimated that $N_{\text{eff}} =$ 113 (105) for the $W^+$ ($W^-$) sample. The high constraining power of the proposed method is manifest in the large reduction of the effective number of replicas.

\begin{figure}
\centerline{
\includegraphics[width=0.5\textwidth]{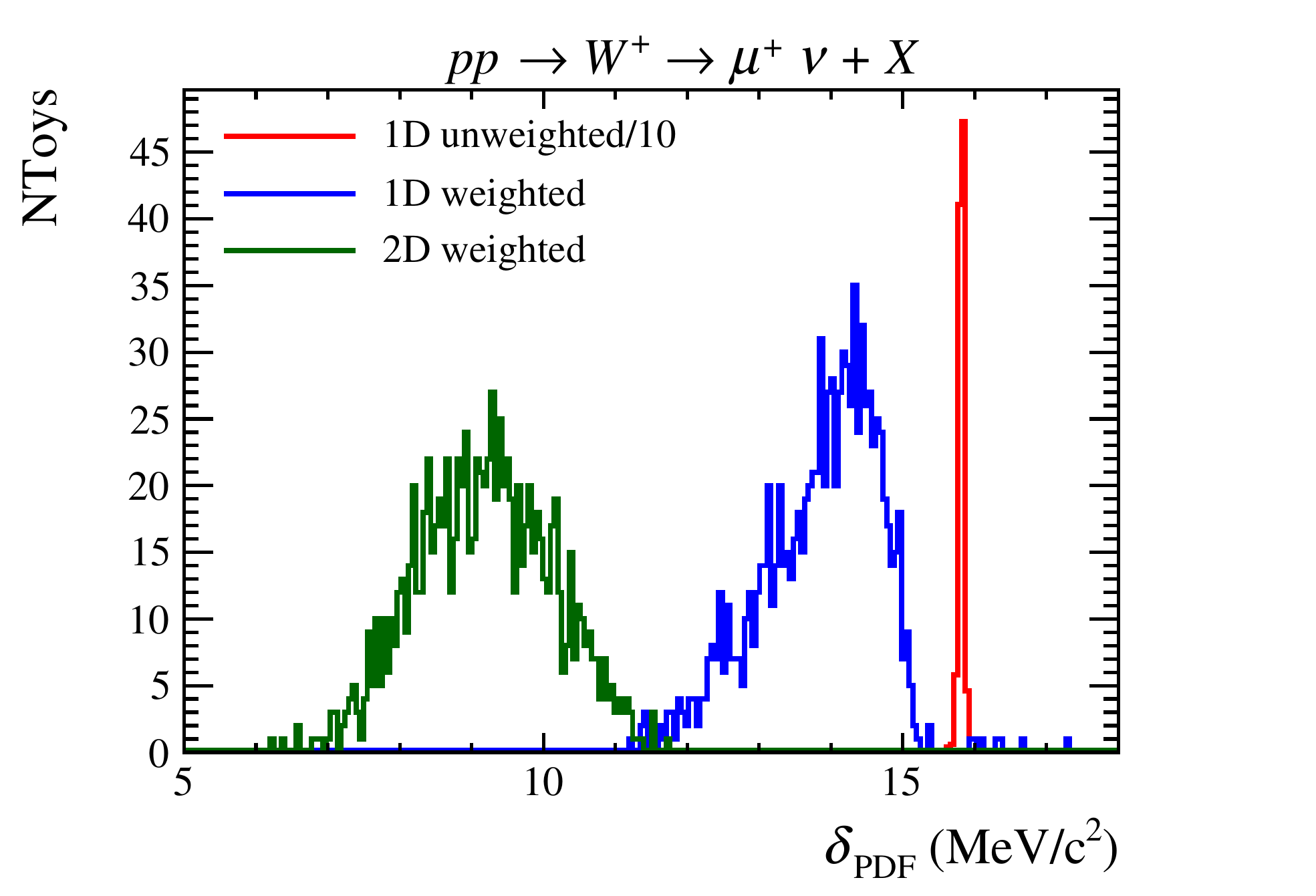}\includegraphics[width=0.5\textwidth]{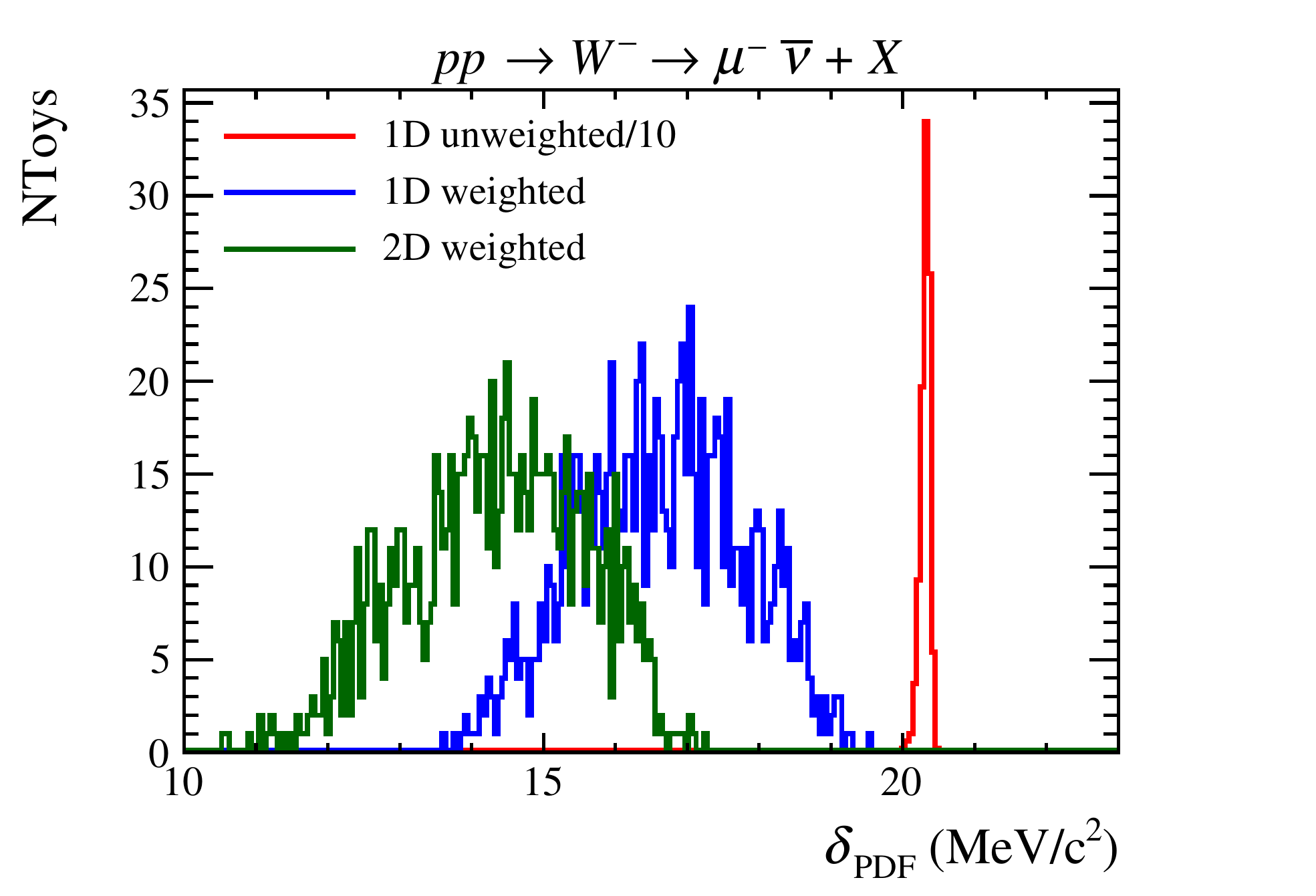}}
\caption{The distribution of the PDF uncertainty evaluated for 1000 toy datasets using three different methods: $p_T^\mu$ fit without weighting, $p_T^\mu$ fit with weighting, ($p_T^\mu$, $\eta$) fit with weighting. The one-dimensional unweighted distribution is arbitrarily scaled down by a factor of ten.}\label{fig:deltaPDF_MultToys}
\end{figure}

The weights are clearly dependent on the toy data, so it is now important to consider the results with multiple toy datasets.
For a single toy dataset the PDF uncertainty is defined by the RMS of the $M_W$ values for the 1000 replicas.
Fig.~\ref{fig:deltaPDF_MultToys} shows the distribution of the PDF uncertainty for 1000 toy datasets,
comparing the one-dimensional fit with and without weights, and the two-dimensional fit with weights.
%In the one-dimensional case the weighting reduces the uncertainty by a typical factor of 10-30\% (10-70\%) for the $W^+$($W^-$).
In the one-dimensional case the weighting reduces the uncertainty by an average factor of 10 (20)\% for the $W^+$ ($W^-$), with a larger spread of the distributions under data fluctuations. In the one-dimensional weighted case this is estimated to be about 0.8 (1.2)\,MeV/c$^2$ for the $W^+$ ($W^-$), in contrast to the 0.04 (0.07)\,MeV/c$^2$ of the unweighted case. 
The two-dimensional weighted case corresponds to a most probable improvement by a factor of roughly two (1.5) for the $W^+$ ($W^-$), with a spread under data fluctuations of 0.9 (1.2)\,MeV/c$^2$. 
Since the outcome of the PDF replica weighting depends on the data, the computation of the PDF uncertainty becomes much more sensitive to the statistical fluctuations of the data themselves. This explains the broadening of the PDF uncertainty distributions once the weighting is applied. This effect becomes even larger for the two-dimensional fit because of its higher constraining power. However, even considering the broadening effect, there is a clear separation between the two-dimensional weighted PDF uncertainty distribution and that of the one-dimensional unweighted (reference) fit approach.
These encouraging results strongly motivate the adoption of the two-dimensional fit method by LHCb.

\subsection{Simultaneous fit of $W^+$ and $W^-$ samples}
\label{sec:WpWm}
\begin{figure}
\centerline{\includegraphics[width=0.5\textwidth]{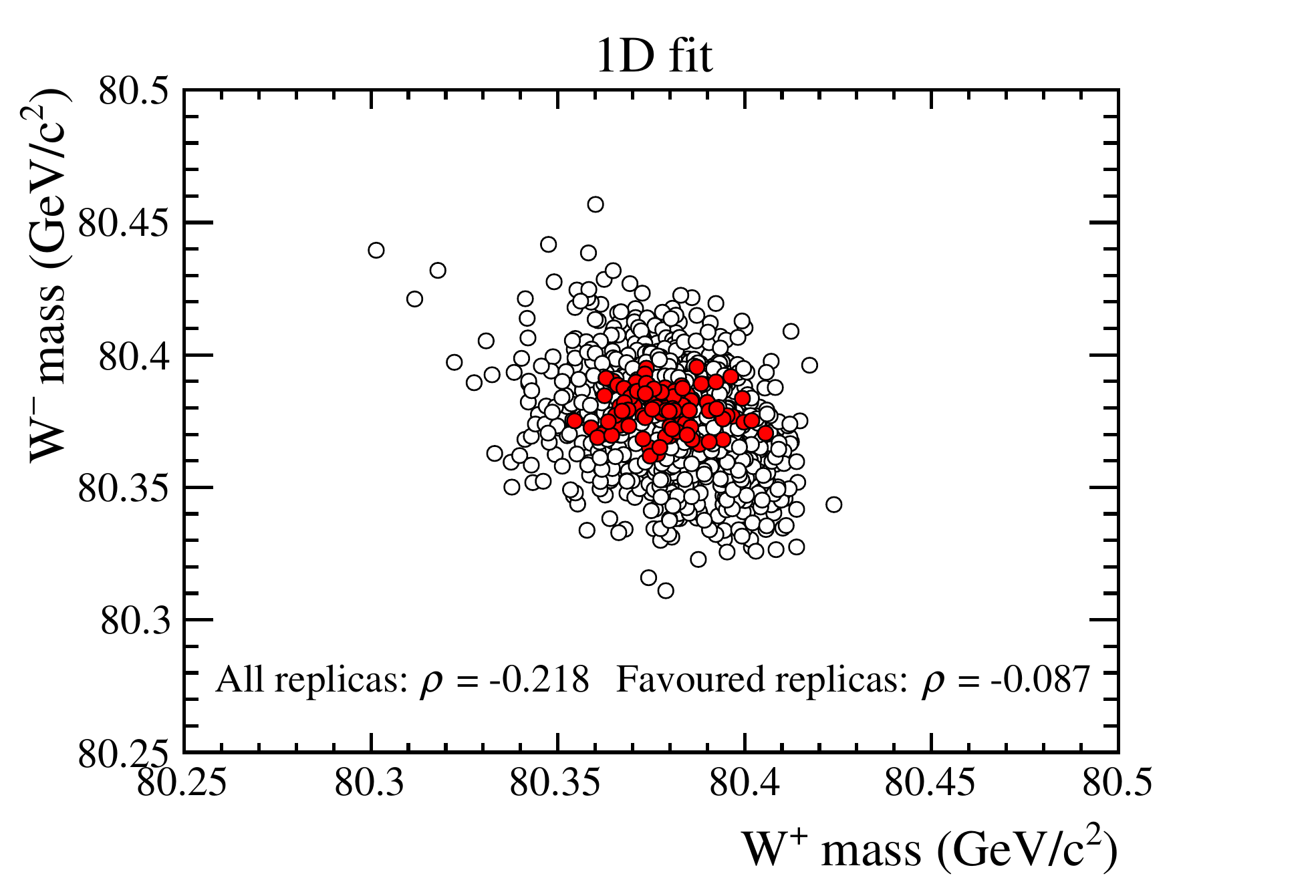}\includegraphics[width=0.5\textwidth]{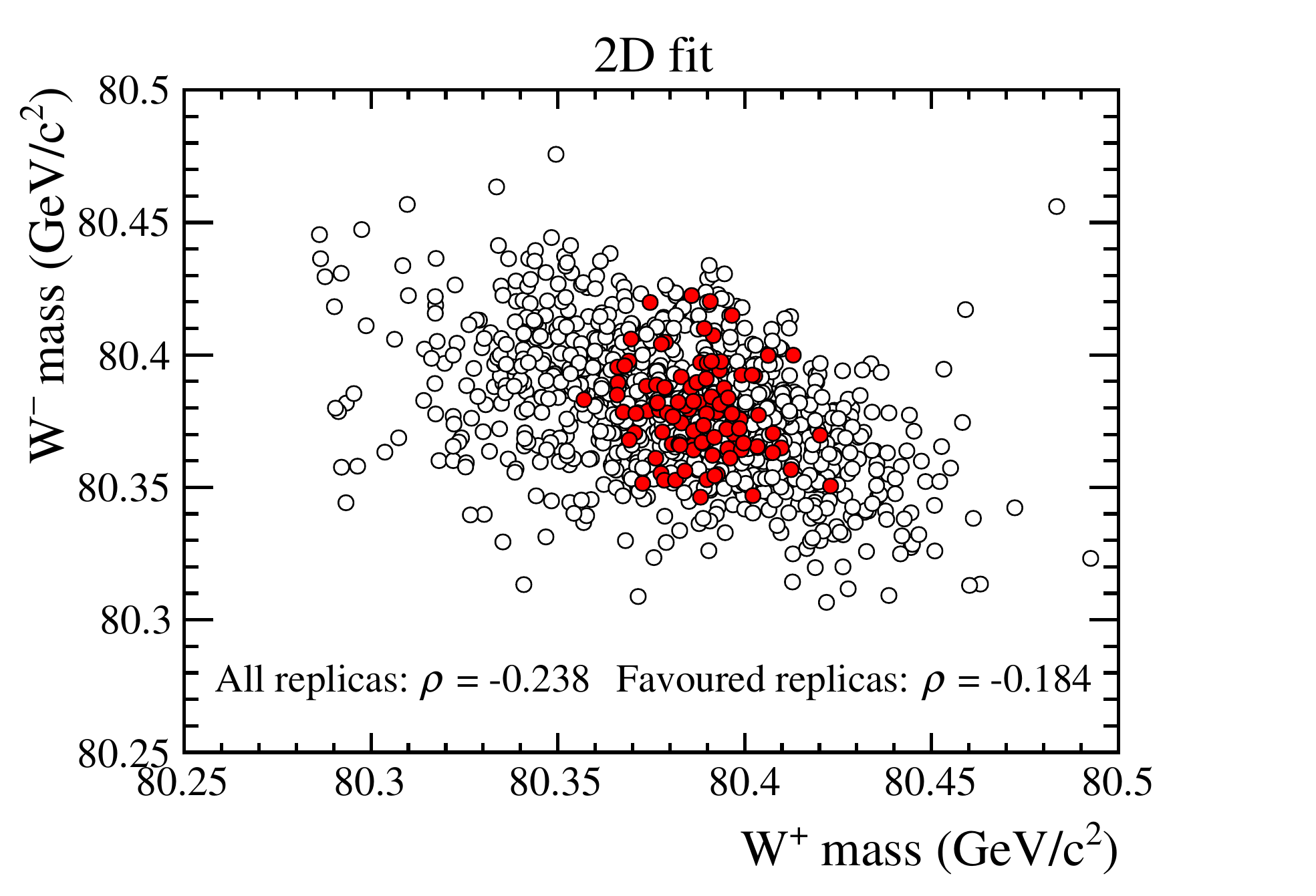}}
\caption{The distribution of the $W^-$ versus $W^+$ mass determined from a single toy dataset with each of the NNPDF3.1 replicas. Ten percent of the replicas with the highest $P(\chi^2)$ (product over the two $W$ charges) are assigned red markers.}\label{fig:Wpm_corr}
\end{figure}

\begin{figure}
\centering
\includegraphics[width=0.5\textwidth]{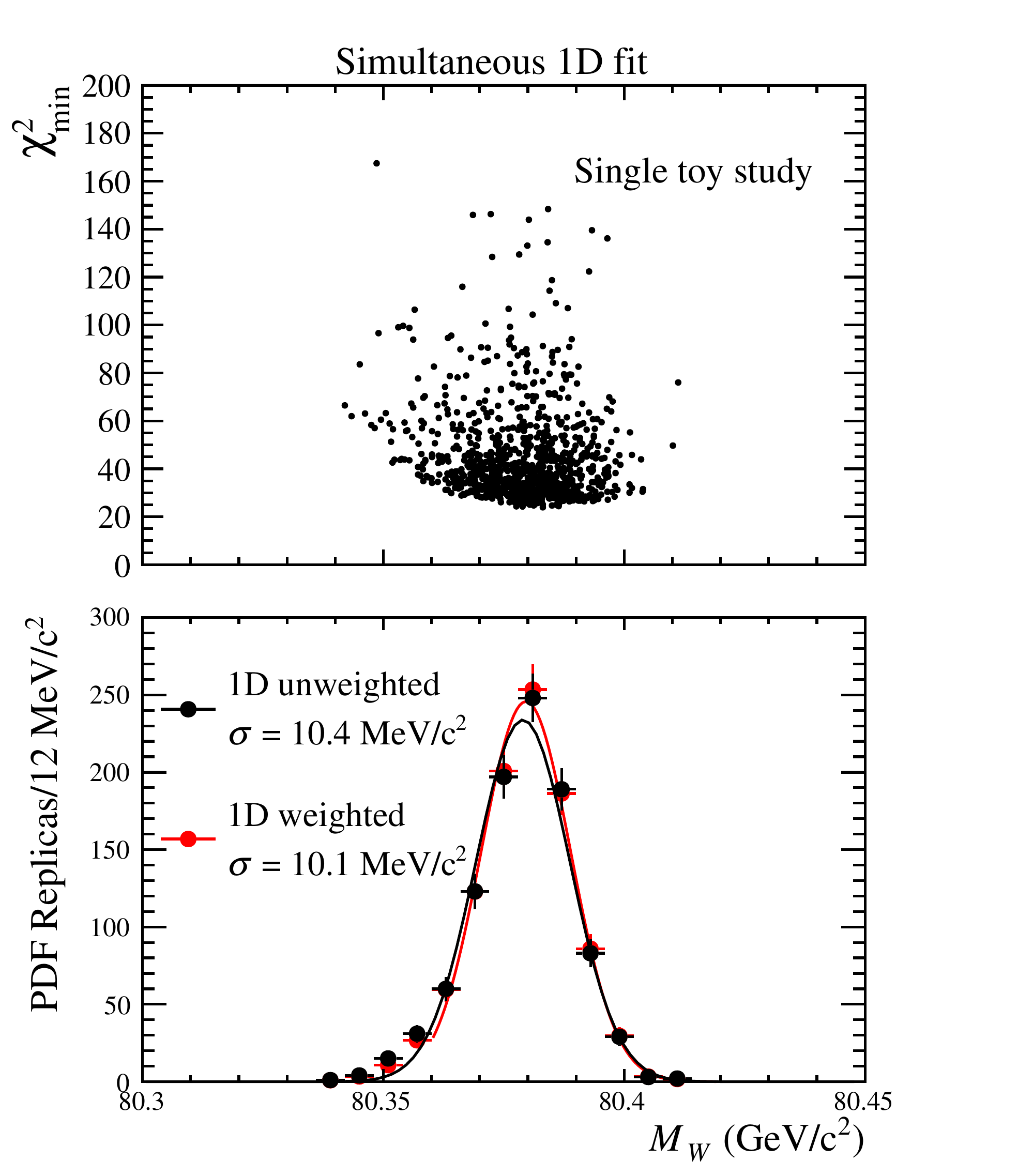}\includegraphics[width=0.5\textwidth]{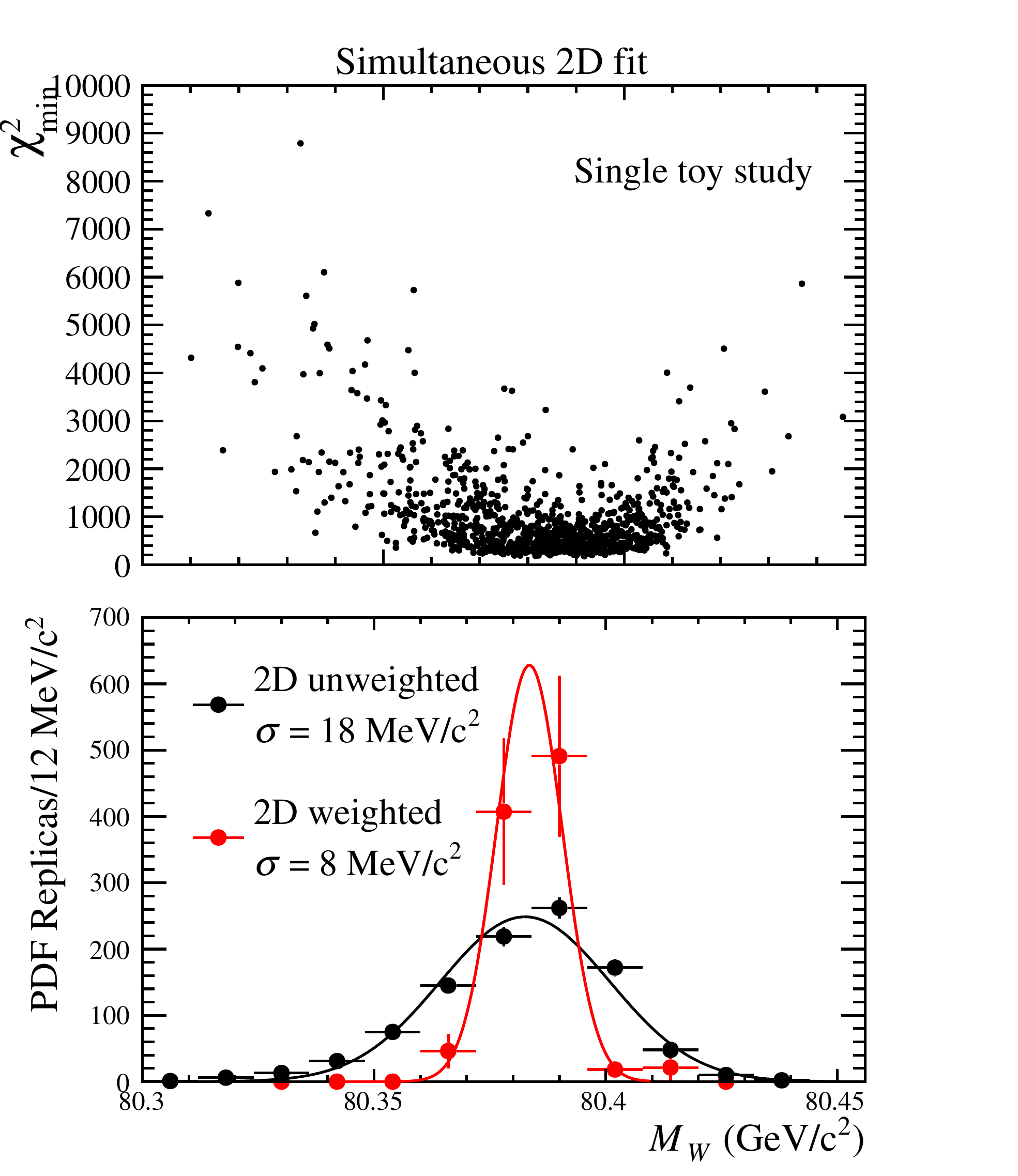}
\caption{Upper: the distribution of $\chi^2$ versus $M_W$ for a one-dimensional (left) and two-dimensional (right) simultaneous fit to a single toy dataset, which assumes the LHCb Run 2 statistics, with each of the 1000 NNPDF3.1 replicas. Lower: the extracted $M_W$ values, with a Gaussian fit function overlaid, without (black) and with (red) weighting. In the simultaneous fit the $W^+$ and $W^-$ templates share the same normalisation.}\label{fig:corr_chi2_mW_SimFit}
\end{figure}

Following the promising results shown for separate fits to the $W^+$ and $W^-$ data it is now interesting to consider the combination of the two charges. Fig.~\ref{fig:Wpm_corr} shows, separately for the one-dimensional and two-dimensional approaches, the $W^+$ versus $W^-$ fit results for a single toy dataset.
Each point represents a different PDF replica. 
Interestingly, for both fit approaches, there is a clear negative correlation, which implies a partial cancellation of the PDF uncertainty when the $W^+$ and $W^-$ data are combined.
It is now interesting to see how this partial anti-correlation is affected by (i) the weights and (ii) moving to a two-dimensional fit.
Therefore, in Fig.~\ref{fig:Wpm_corr} ten percent of the points corresponding to the largest product (over the two $W$ charges) of $P(\chi^2_{\text{min}})$ values are highlighted.
Unfortunately, in both the one- and two-dimensional fit cases, the subset of favoured replicas exhibits a correlation coefficient with a reduced magnitude.
%Before attempting to perform a fit to the combined $W^+$ and $W^-$ data there is a further consideration.
%Fig.~\ref{fig:muETA_chargeasy} shows the variation between the replicas of the charge asymmetry between the $W^+$ and $W^-$ cross sections, as a function of $\eta$.
%The lines are assigned a colour according to the shift in the $M_W$ value determined in a simultaneous fit (one-dimensional) of the $W^+$ and $W^-$ data.
%A clear pattern can been seen whereby replicas that bias $M_W$ upwards(downwards) tend to shift the charge asymmetry upwards(downwards), particularly at large $\eta$ values.
%Most of the $\eta$ dependence is already taken into account with the weighted two-dimensional fit, but the integrated charge asymmetry can
%be trivially included as a constraint by sharing the normalisation factor between the two $W$ charges.
Fig.~\ref{fig:corr_chi2_mW_SimFit} shows the $\chi^2_{\text{min}}$ versus $M_W$ values for combined ($W^+$ and $W^-$) fits to a single toy dataset. The normalisation for both the datasets is scaled by the same parameter to take into account the integrated charge asymmetry constraint on the PDFs. Each point corresponds to a different NNPDF3.1 replica,
and the results are shown separately for the one-dimensional and two-dimensional fits.
The weighted and unweighted $M_W$ distributions are shown with corresponding Gaussian fits overlaid.
With these data the weights have very little effect on the width of the distribution in the one-dimensional case. The effective number of replicas ($N_{\text{eff}}$) after reweighting, computed using Eq.~\ref{eq:Neff}, is indeed 928. 
In the two-dimensional case, however, there is roughly a factor of two of improvement. The effective number of replicas estimated for this case ($N_{\text{eff}}$ = 35) is showing a very large constraining power of the data and suggests that, for the final measurement, a more robust approach like the Hessian method or an increase of the number of replicas in the reweigting procedure, is necessary to guarantee the statistical reliability of the results obtained with the two-dimensional fit. 

\begin{figure}
\centerline{\includegraphics[width=0.5\textwidth]{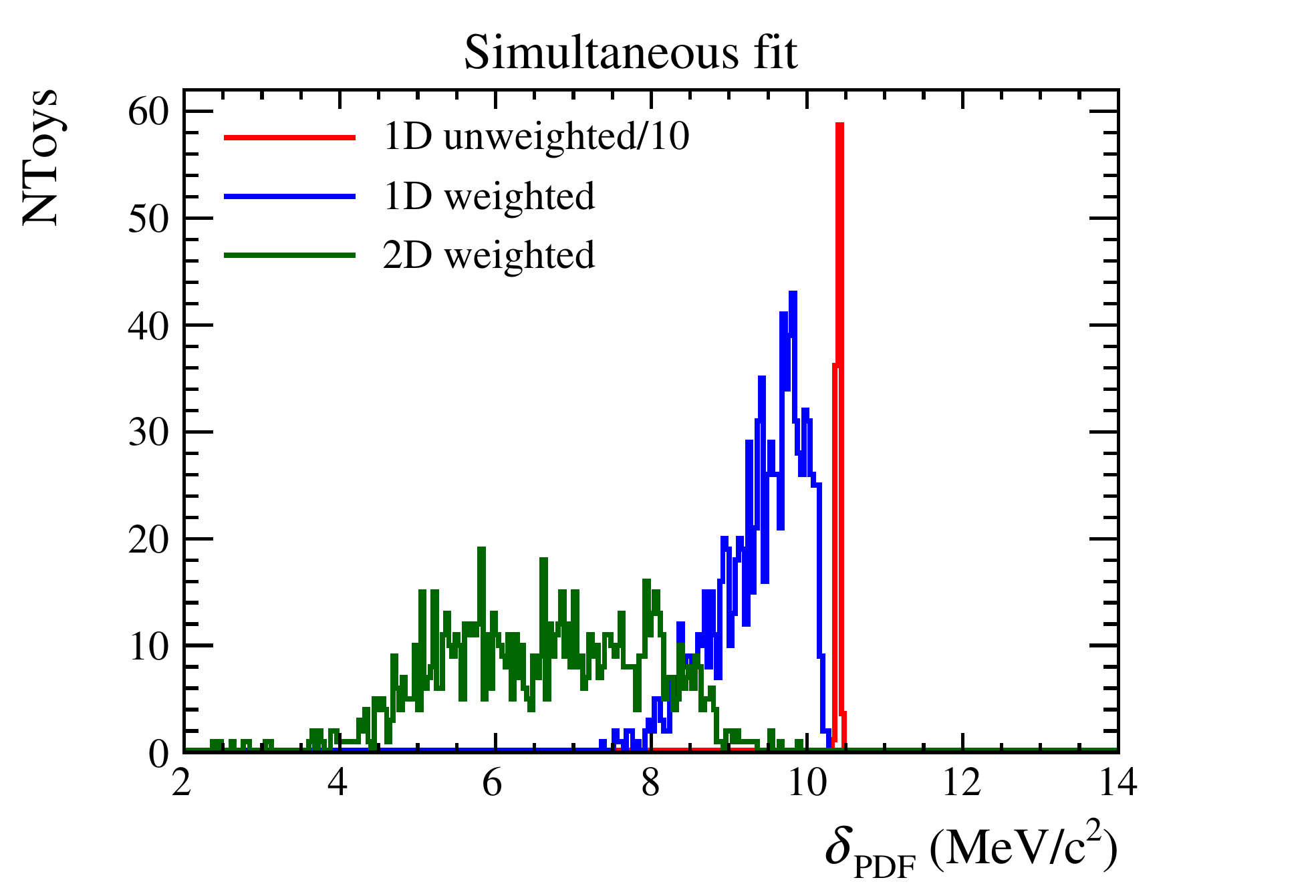}\includegraphics[width=0.5\textwidth]{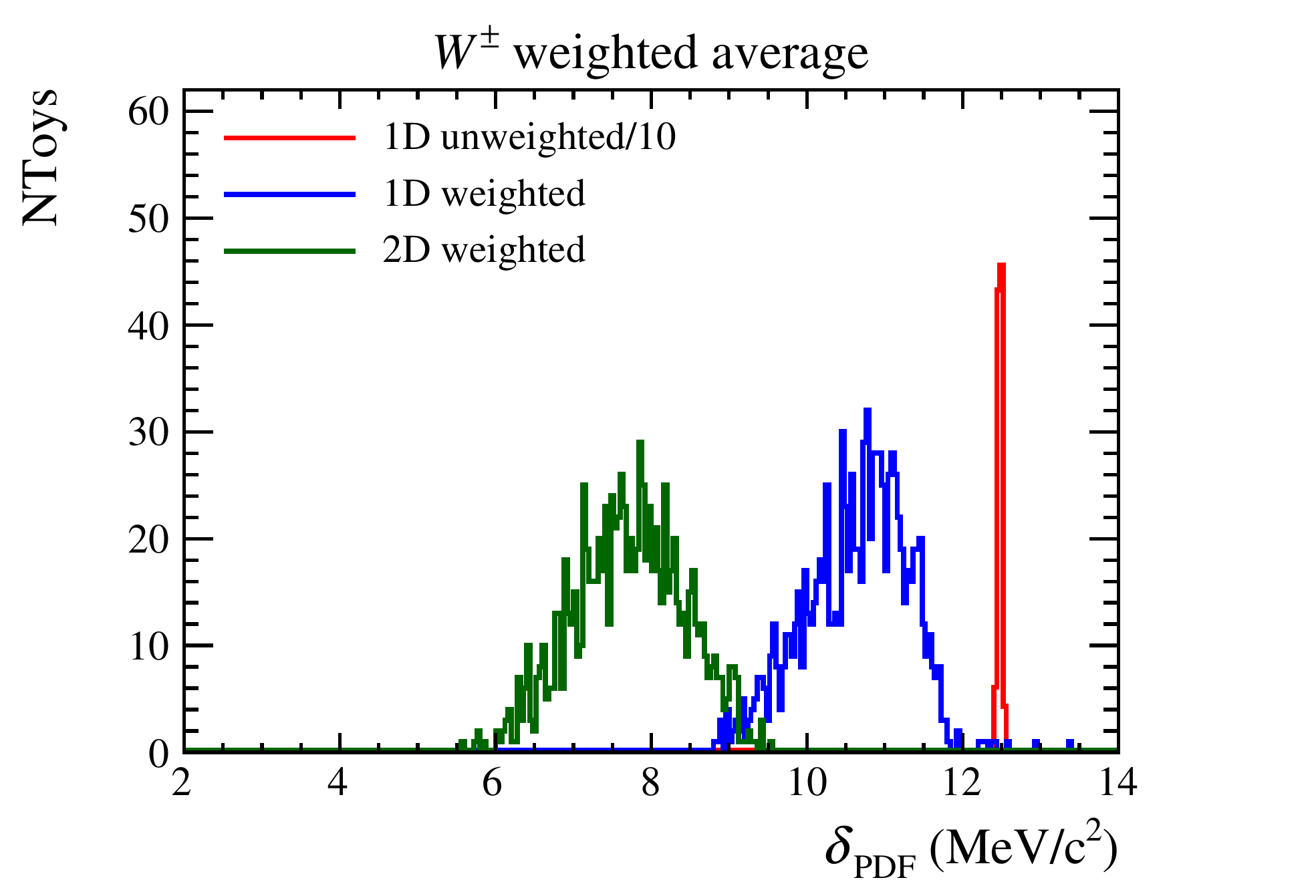}}
\caption{The distribution of the PDF uncertainty evaluated for 1000 toy datasets from (left) a simultaneous fit where the $W^+$ and $W^-$ templates share the same normalisation or (right) a weighted average of single $W$ charges measurements. 
Three different fit methods are compared: $p_T^\mu$ fit without weighting, $p_T^\mu$ fit with weighting, ($p_T^\mu$, $\eta$) fit with weighting. The one-dimensional unweighted distribution is arbitrarily scaled down by a factor of ten.}\label{fig:deltaPDF_MultToys_SimFit}
\end{figure}

Fig.~\ref{fig:deltaPDF_MultToys_SimFit} (left) shows the distribution of the PDF uncertainty in 1000 toy datasets, in combined fits of the $W^+$ and $W^-$ data.
Compared to the traditional one-dimensional fit, the addition of the weighting typically improves the PDF uncertainty by around 10\%.
The two-dimensional fit with weighting is, however, typically around a factor of two better.
If the normalisation is no longer shared between the $W^+$ and $W^-$ the uncertainty is typically slightly larger, but this change is usually less than 1 MeV/c$^2$.
Fig.~\ref{fig:deltaPDF_MultToys_SimFit}  (right) considers an alternative approach whereby the $W^+$ and $W^-$ data are analysed separately, and the corresponding $M_W$ values are combined in a weighted average.
This results in larger uncertainties, and therefore encourages the simultaneous fit of $W^+$ and $W^-$ data with a single shared $M_W$ fit parameter.
%\begin{figure}
%\centerline{\includegraphics[width=0.5\textwidth]{figures_tempPDF/muETA_ChargeAsy_asymm_mWweight.pdf}}
%\caption{The charge asymmetry as a function of $\eta$ for the first 100 NNPDF3.1 replicas. The colour scale indicates the shift of the $M_W$ value extracted for each replica with respect to the central replica using a simultaneous one-dimensional fit of $W^+$ and $W^-$ samples.}\label{fig:muETA_chargeasy}
%\end{figure}

\subsection{Dependence on the detector acceptance}
\label{sec:fitrange}
\begin{figure}
\centering
\includegraphics[width=0.5\textwidth]{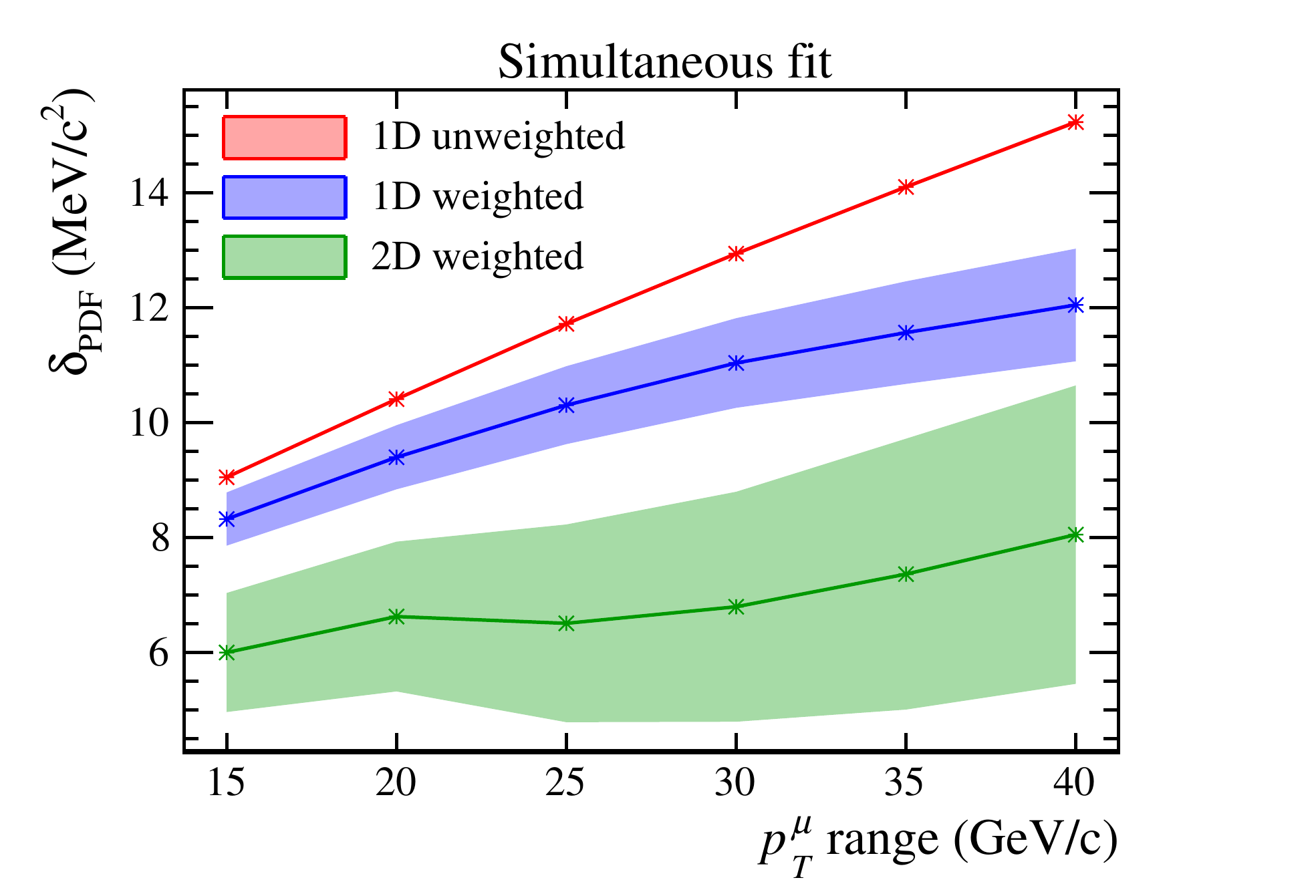}
\caption{PDF uncertainty as a function of the $p_T^\mu$ range (the full width, centered around $M_W/2$) used in the simultaneous fit. The bands report the mean and the RMS of the distribution of the PDF uncertainty evaluated for 1000 toy datasets. The $\eta$ range is set to 2 $<\eta<$ 4.5. In the two dimensional fits three $\eta$ bins are used.}\label{fig:FitRange_PTbin}
\end{figure}

\begin{figure}
\centering
\includegraphics[width=0.5\textwidth]{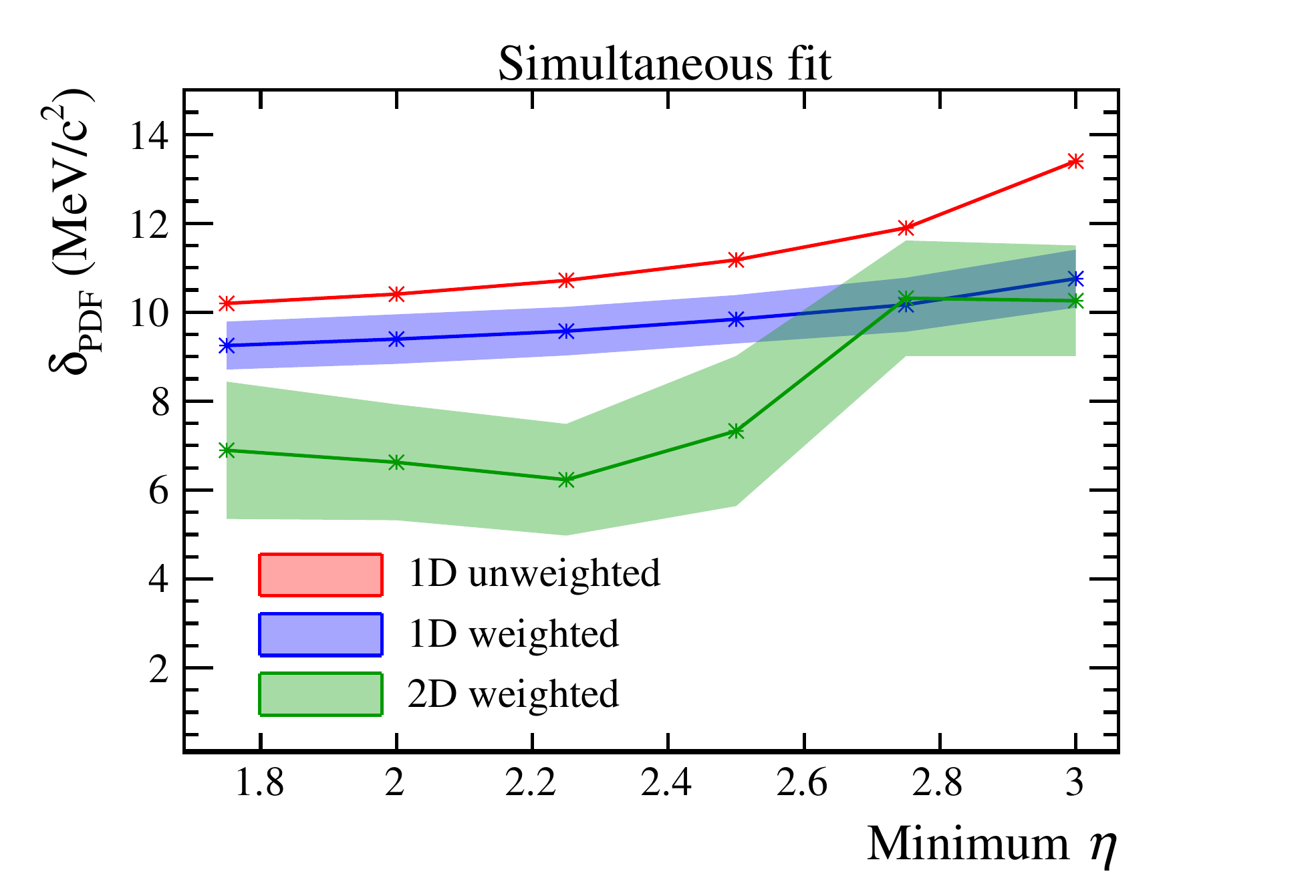}\includegraphics[width=0.5\textwidth]{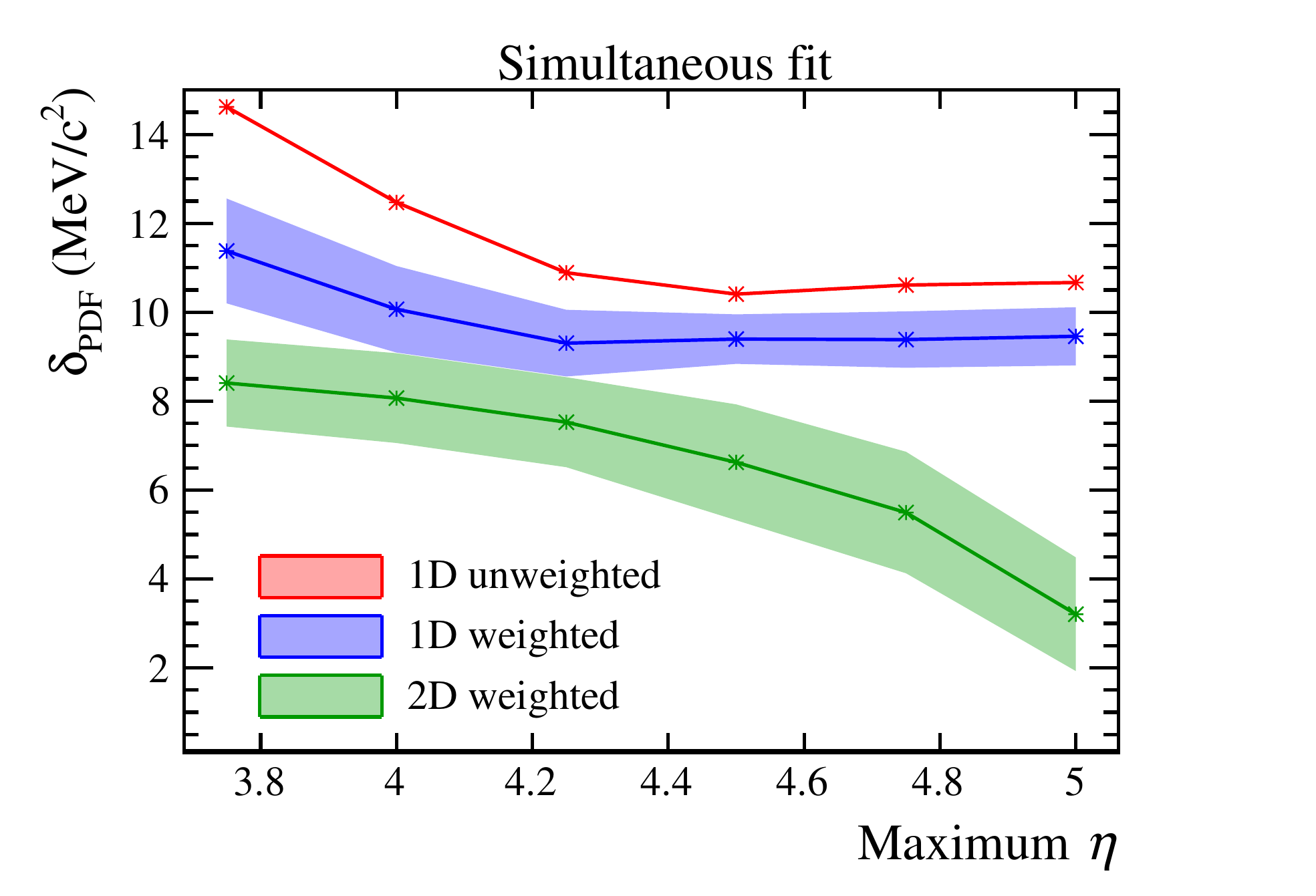}
\caption{PDF uncertainty as a function of the lower (left) and upper (right) $\eta$ cut used in the simultaneous fit. The bands report the mean and the RMS of the distribution of the PDF uncertainty evaluated for 1000 toy datasets. The $p_T^\mu$ is in the range 30 $<p_T^\mu<$ 50\,GeV/c. Left (Right): the upper (lower) $\eta$ cut set to 4.5 (2). In the two dimensional fits three $\eta$ bins are used.}\label{fig:sigmaPDF_etacut}
\end{figure}

The study has thus far restricted to events in the range $30 < p_T^\mu < 50$~GeV/c and $2 < \eta < 4.5$. 
It is interesting to now consider how the results depend on this choice, since the LHCb acceptance extends slightly outside this eta range, and LHCb is able to trigger on muons with far smaller $p_T^\mu$ values without any prescales.
Fig.~\ref{fig:FitRange_PTbin} shows how the PDF uncertainties depend on the width of the $p_T^\mu$ interval,
which is symmetric around $M_W/2$. Each band is centered on the mean of the distribution of the PDF uncertainty evaluated for 1000 toy datasets and its width is defined as the RMS of the same distribution. 
With the simple one-dimensional unweighted fit the PDF uncertainty grows approximately linearly with the width of the $p_T^\mu$ interval. 
This is also the case for the one- and two-dimensional weighted fits, though the slope is less severe. Despite this study suggests that choosing a smaller fit range yields to smaller PDF uncertainties, the reduction of this range has an impact on the statistical precision of the measurement as well. 
Fig.~\ref{fig:sigmaPDF_etacut} considers separately the dependence on the minimum and maximum $\eta$ value.
The uncertainty is found to reduce when the $\eta$ range is extended in either direction. 
The uncertainty is not significantly changed if the number of $\eta$ bins is increased from the nominal value of three.
Using only three bins in $\eta$ should make the experimental control of the $\eta$ dependence of the muon efficiency more straightforward to control than if more bins are required.
\section{Conclusions}
\label{sec:conclusions}
It has recently been suggested that LHCb should perform a measurement of $M_W$ based on a one-dimensional fit to the
muon $p_T^\mu$ distribution in samples of $W \to \mu\nu$ decays.
Thanks to LHCb's unique angular coverage this measurement would complement those performed by ATLAS and CMS, particularly when considering PDF uncertainties.
Here we report on a detailed study of the PDF uncertainty, restricting to the NNPDF3.1 set, on the proposed LHCb measurement.
It is found that the variations in the PDFs that tend to bias the determination of $M_W$ lead to clear patterns
of variation in the shapes of the $W$ kinematic distributions, in particular the rapidity distribution.
A particularly interesting observation is that those variations also lead to a {\em measurable} change in the shape of the muon $\eta$ distribution.
%Given that the PDF variations also fail to perfectly mimic a change in $M_W$ in the $p_T^\mu$ spectrum, 
%this suggests that a two-dimensional fit to the $p_T^\mu$ versus $\eta$ distribution, with PDF replica weighting, would allow the PDF uncertainty to be constrained.
An analysis performed on a two-dimensional ($p_T^\mu$ versus $\eta$) plane would reduce the capability of the PDFs to give rise to changes in the $p_T^\mu$ distribution that can be misidentified as variations of $M_W$. Therefore, with large enough data samples, a two-dimensional fit to the $p_T^\mu$ versus $\eta$ distribution, with PDF replica weighting, would allow the PDF uncertainty to be further constrained. 
A study with 1000 experiments, assuming the LHCb Run 2 statistics, indicates a typical improvement of around a factor of two, compared to the one-dimensional fit to the $p_T^\mu$ spectrum alone, when fitting the $W^+$ and $W^-$ data simultaneously. 
Alternative approaches to the PDF replica reweighting, such as the Hessian method, should be considered in future studies towards the real measurement. 
The full PDF uncertainty should also include the variation between results from different PDF fitting groups, but this is a very encouraging result.
In order to facilitate the study of the possible impact of other data a table of $M_W$ biases for the first 100 NNPDF3.1 replicas is provided as supplementary material.
The main study considers events in which the muon satisfies $2 < \eta < 4.5$ and $30 < p_T^\mu < 50$\,GeV/c, but the dependence on these choices is also studied 
since there are likely to be many considerations on the optimal fit range for the real measurement.
\section*{Acknowledgements}
\noindent We thank W.~Barter, M.~Charles, G.~Bozzi, A.~Vicini, A.~Cooper-Sakar,  L.~Harland-Lang and J.~Rojo for their helpful comments
and suggestions during the preparation of this manuscript.
OL thanks the CERN LBD group for their support during the period when most of this work was carried out,
and MV thanks the Science and Technologies Facilities Council for their support through an Ernest Rutherford Fellowship.

\addcontentsline{toc}{section}{References}
\bibliographystyle{LHCb}
\bibliography{mybib}
 
\addcontentsline{toc}{section}{Additional material}
\newpage
\thispagestyle{empty}
\section*{Additional material}
\label{sec:additional}

\begin{table}[H]
\begin{center}
\resizebox*{!}{0.9\textheight}{
\begin{tabular}{| c| c | c| c|  c |c|c| }
  \hline
 & \multicolumn{2}{c}{$\Delta M_+$ (GeV/c$^2$)} & \multicolumn{2}{|c|}{$\Delta M_-$ (GeV/c$^2$)} &  \multicolumn{2}{|c|}{$\Delta M_\pm$ (GeV/c$^2$)}\\
  \hline
  $R_{\text{NNPDF3.1}}$  & 1D fit & 2D fit & 1D fit & 2D fit & 1D fit  & 2D fit\\
  \hline		 
 1 & 0.004 & 0.0191 & 0.0159 & 0.0134 & 0.008 & 0.0172\\
2 & 0.0195 & 0.0169 & -0.0349 & -0.0378 & 0.0007 & -0.0011\\
3 & -0.0074 & -0.0247 & -0.0063 & 0.0012 & -0.007 & -0.0163\\
4 & 0.0071 & -0.0015 & 0.0392 & 0.0438 & 0.0181 & 0.0133\\
5 & -0.0068 & -0.0057 & -0.0164 & -0.0141 & -0.0101 & -0.0085\\
6 & 0.0032 & -0.0043 & 0.0053 & 0.0077 & 0.0039 & -0.0004\\
7 & 0.0049 & -0.0051 & -0.0048 & -0.0087 & 0.0015 & -0.0064\\
8 & 0.011 & 0.0174 & -0.0119 & -0.0183 & 0.0031 & 0.0057\\
9 & 0.0057 & 0.0392 & 0.0072 & -0.0048 & 0.0062 & 0.0248\\
10 & -0.0077 & -0.0402 & -0.0015 & 0.0019 & -0.0056 & -0.0264\\
11 & 0.0077 & 0.0248 & -0.0158 & -0.0213 & -0.0004 & 0.0096\\
12 & -0.0168 & -0.0353 & -0.0076 & -0.0107 & -0.0137 & -0.0273\\
13 & -0.012 & -0.014 & -0.0134 & -0.0117 & -0.0125 & -0.0133\\
14 & 0.0087 & 0.0165 & -0.0197 & -0.028 & -0.0011 & 0.0018\\
15 & -0.0308 & -0.0611 & -0.0133 & -0.0141 & -0.0248 & -0.0457\\
16 & -0.0063 & -0.0268 & 0.0152 & 0.0247 & 0.001 & -0.01\\
17 & -0.0103 & -0.0359 & -0.0033 & 0.4204 & -0.0079 & -0.0229\\
18 & 0.02 & 0.0266 & -0.0307 & -0.0467 & 0.0025 & 0.0024\\
19 & 0.0037 & -0.015 & -0.0106 & -0.0046 & -0.0012 & -0.0116\\
20 & -0.0002 & -0.0045 & -0.0382 & -0.0405 & -0.0133 & -0.0164\\
21 & -0.029 & -0.0688 & 0.0082 & 0.0226 & -0.0162 & -0.0389\\
22 & 0.0166 & 0.0233 & 0.0011 & -0.0003 & 0.0112 & 0.0155\\
23 & 0.0204 & 0.0333 & -0.0287 & -0.0364 & 0.0035 & 0.0104\\
24 & 0.0185 & 0.0279 & -0.021 & -0.0313 & 0.005 & 0.0084\\
25 & 0.0027 & 0.0287 & 0.0045 & -0.0054 & 0.0033 & 0.0174\\
26 & -0.0036 & 0.007 & 0.0331 & 0.0492 & 0.009 & 0.0208\\
27 & -0.0191 & -0.0289 & -0.0195 & -0.0188 & -0.0193 & -0.0256\\
28 & -0.0056 & 0.008 & -0.0082 & -0.0161 & -0.0065 & -0.0073\\
29 & 0.016 & 0.012 & 0.0221 & 0.4204 & 0.0076 & 0.0156\\
30 & 0.0149 & 0.036 & 0.0147 & 0.0106 & 0.4214 & 0.418\\
31 & 0.0246 & 0.0389 & -0.0305 & -0.0375 & 0.0057 & 0.0138\\
32 & -0.0101 & -0.0299 & 0.0078 & 0.0152 & -0.004 & -0.0152\\
33 & 0.0304 & 0.0531 & -0.0167 & -0.0323 & 0.0141 & 0.0248\\
34 & -0.0255 & -0.0445 & 0.0231 & 0.0329 & -0.0088 & -0.0191\\
35 & -0.0214 & -0.0307 & 0.0053 & -0.0038 & -0.0122 & -0.0219\\
36 & -0.0021 & -0.0043 & 0.0404 & 0.0467 & 0.0125 & 0.0123\\
37 & 0.0324 & 0.0365 & -0.0085 & -0.0079 & 0.0183 & 0.0219\\
38 & -0.0054 & -0.0005 & -0.0139 & -0.0122 & -0.0084 & -0.0044\\
39 & 0.0141 & 0.0433 & -0.0321 & -0.0475 & -0.0018 & 0.0132\\
40 & -0.0327 & -0.0576 & -0.0052 & -0.0203 & -0.0232 & -0.0454\\
41 & -0.0171 & -0.0141 & 0.0175 & 0.0181 & -0.0052 & -0.0036\\
42 & -0.0142 & -0.0189 & 0.0127 & 0.0147 & -0.005 & -0.0079\\
43 & 0.0061 & 0.0172 & -0.0212 & -0.0281 & -0.0033 & 0.0022\\
44 & 0.0095 & 0.018 & -0.0218 & -0.0291 & -0.0012 & 0.0025\\
45 & 0.0055 & 0.0025 & 0.0064 & 0.0081 & 0.0058 & 0.0043\\
46 & -0.0262 & -0.0478 & 0.0297 & 0.0429 & -0.007 & -0.0182\\
47 & -0.0121 & -0.0528 & -0.0415 & -0.0398 & -0.0224 & -0.0486\\
48 & -0.018 & -0.028 & 0.0129 & 0.0234 & 0.0044 & -0.0113\\
49 & 0.0021 & -0.0017 & -0.0131 & -0.0144 & -0.0032 & -0.0059\\
50 & 0.0206 & 0.0618 & -0.0353 & -0.0511 & 0.0014 & 0.0244\\
51 & 0.0067 & 0.0151 & -0.0086 & -0.0126 & 0.0015 & 0.0059\\
52 & -0.0155 & -0.0152 & 0.0149 & 0.0163 & -0.005 & -0.0049\\
53 & -0.0035 & 0.0127 & 0.0288 & 0.0316 & 0.0076 & 0.0189\\
54 & -0.0254 & -0.0514 & -0.0278 & -0.016 & -0.0263 & -0.0399\\
55 & 0.0055 & 0.0068 & -0.0238 & -0.0267 & -0.0046 & -0.0043\\
56 & 0.0032 & 0.0401 & 0.0035 & -0.0028 & 0.0033 & 0.026\\
57 & 0.0129 & 0.0237 & 0.0217 & 0.0212 & 0.016 & 0.0228\\
58 & 0.0035 & 0.006 & 0.0155 & 0.0185 & 0.0076 & 0.0101\\
59 & -0.0095 & -0.0232 & 0.0131 & 0.0219 & -0.0018 & -0.0085\\
60 & 0.0156 & 0.0243 & -0.0228 & -0.0259 & 0.0024 & 0.0077\\
61 & 0.0281 & 0.0312 & 0.012 & 0.0055 & 0.0225 & 0.0227\\
62 & -0.0065 & -0.0032 & 0.0098 & 0.0018 & -0.1034 & -0.1068\\
63 & 0.0074 & 0.0155 & 0.013 & 0.0097 & 0.0093 & 0.0135\\
64 & 0.0283 & 0.0771 & -0.0492 & -0.0666 & 0.0016 & 0.0294\\
65 & 0.0042 & -0.0175 & -0.0252 & -0.0297 & -0.0059 & -0.0215\\
66 & -0.0004 & -0.0015 & 0.0364 & 0.0427 & 0.0122 & 0.0129\\
67 & 0.0014 & 0.0235 & 0.0249 & 0.0154 & 0.0094 & 0.0208\\
68 & 0.0135 & 0.0357 & -0.0249 & -0.0352 & 0.0003 & 0.0123\\
69 & 0.0239 & 0.0523 & -0.0081 & -0.0206 & 0.0129 & 0.0283\\
70 & 0.0134 & 0.0304 & -0.0225 & -0.0264 & 0.001 & 0.0116\\
71 & 0.0073 & 0.0062 & -0.012 & -0.0226 & 0.0006 & -0.0033\\
72 & 0.0057 & 0.0055 & -0.0168 & -0.0126 & -0.0021 & -0.0005\\
73 & 0.0073 & 0.0051 & -0.0321 & -0.0333 & -0.0063 & -0.0076\\
74 & 0.0055 & 0.0009 & -0.0203 & -0.0234 & -0.0034 & -0.0071\\
75 & 0.0139 & 0.0102 & 0.0115 & 0.0146 & 0.0131 & 0.0116\\
76 & -0.0091 & -0.0482 & -0.0133 & 0.0021 & -0.0105 & -0.0317\\
77 & 0.0045 & 0.0161 & 0.0095 & 0.0094 & 0.0062 & 0.0138\\
78 & -0.0054 & -0.0064 & 0.0024 & 0.0024 & -0.0028 & -0.0036\\
79 & 0.0015 & 0.0008 & 0.0095 & 0.0025 & 0.0042 & 0.0013\\
80 & -0.0274 & -0.0431 & 0.0114 & 0.0111 & -0.0141 & -0.0253\\
 81 & -0.0018 & -0.0217 & -0.0147 & 0.0018 & -0.0063 & -0.014\\
82 & -0.0145 & -0.0203 & 0.0227 & 0.0271 & -0.0017 & -0.0048\\
83 & -0.0001 & -0.0404 & 0.0205 & 0.0301 & 0.007 & -0.0173\\
84 & -0.0088 & -0.0383 & 0.0106 & 0.0312 & -0.0021 & -0.0155\\
85 & -0.0276 & -0.0562 & -0.0062 & -0.0035 & -0.0203 & -0.039\\
86 & -0.0094 & -0.0184 & -0.003 & 0.0076 & -0.0072 & -0.0099\\
87 & -0.0032 & 0.0105 & 0.0204 & 0.0173 & 0.0049 & 0.0127\\
88 & 0.0001 & -0.0057 & -0.0042 & -0.0023 & -0.0014 & -0.0046\\
89 & 0.001 & 0.036 & 0.0235 & 0.0128 & 0.0087 & 0.0283\\
90 & -0.0091 & -0.0416 & -0.0012 & 0.0066 & -0.0064 & -0.0258\\
91 & 0.0267 & 0.0642 & -0.005 & -0.0197 & 0.0158 & 0.0364\\
92 & -0.0081 & -0.0485 & 0.0011 & 0.0125 & -0.0049 & -0.0286\\
93 & -0.0037 & -0.0077 & 0.0214 & 0.0284 & 0.0049 & 0.0041\\
94 & 0.0083 & 0.0066 & 0.0158 & 0.0182 & 0.0108 & 0.0104\\
95 & 0.0052 & 0.0144 & -0.006 & -0.0069 & 0.0014 & 0.0073\\
96 & -0.0026 & -0.0041 & -0.0457 & -0.0504 & -0.0174 & -0.0194\\
97 & 0.0038 & 0.0209 & -0.0088 & -0.0125 & -0.0006 & 0.0098\\
98 & -0.0111 & -0.0141 & 0.009 & 0.0169 & -0.0042 & -0.004\\
99 & 0.0025 & 0.0046 & 0.0309 & 0.0352 & 0.0122 & 0.0146\\
100 & 0.0082 & 0.0224 & -0.0132 & -0.0129 & 0.0008 & 0.0108\\
\hline
\end{tabular}
}% resizebox
\caption{Shift in the extracted $M_W$ values for the first 100 NNPDF3.1 replicas with respect to the central replica. Both single $W$ charges results and the simultaneous fit results are shown. The $\Delta M $ values are extracted from both the one-dimensional and the two-dimensional fit. In the simultaneous fit the normalisation constraint of the $W^+$ and $W^-$ templates is taken into account.}\label{tab:shifts_mW}
\end{center}
\end{table}

\end{document}